\documentclass[journal]{IEEEtran}

\usepackage[numbers]{natbib}

\usepackage{xcolor,soul,framed} 

\colorlet{shadecolor}{yellow}
\usepackage[pdftex]{graphicx}
\graphicspath{{../pdf/}{../jpeg/}}
\DeclareGraphicsExtensions{.pdf,.jpeg,.png}

\usepackage[cmex10]{amsmath}
\usepackage{array}
\usepackage{mdwmath}
\usepackage{mdwtab}
\usepackage{eqparbox}
\usepackage{url}
\usepackage{amsfonts}
\usepackage{cleveref}
\usepackage{subfig}
\usepackage[scr=rsfs]{mathalpha}
\usepackage{makecell}
\usepackage{float} 
\usepackage[printonlyused]{acronym}

\crefalias{subequation}{equation} 
\crefalias{eqnarray}{equation} 
\crefformat{pluraleq}{(#2#1#3)}

\usepackage{bm}
\usepackage[ruled]{algorithm2e}

\newcommand{\be}{\boldsymbol{e}}

\newcommand{\bn}{\boldsymbol{n}}

\newcommand{\bt}{\boldsymbol{t}}

\newcommand{\bx}{\boldsymbol{x}}

\begin{document}
	\bstctlcite{IEEEexample:BSTcontrol}
	\title{A data-enabled physics-informed neural network with comprehensive numerical study on solving neutron diffusion eigenvalue problems}
	\author{Yu Yang, Helin Gong*, Shiquan Zhang*, Qihong Yang, Zhang Chen, Qiaolin He, Qing Li 
		\thanks{Corresponding author: Helin Gong (gonghelin@sjtu.edu.cn), Shiquan Zhang (shiquanzhang@scu.edu.cn)}
		\thanks{Helin Gong is with ParisTech Elite Institute of Technology, Shanghai Jiao Tong University, 200240, Shanghai, China.}
		\thanks{Yu Yang, Shiquan Zhang, Qihong Yang and Qiaolin He are with School of Mathematics, Sichuan University, 610065, Chengdu, China.}
		\thanks{Zhang Chen and Qing Li are with Science and Technology on Reactor System Design Technology Laboratory, Nuclear Power Institute of China, 610041, Chengdu, China.}
	} 
	
	
	\markboth{Annals of Nuclear Energy, VOL. XX, NO. XX, XXXX
		2022}{Yang {\textit{et al.}}: A comprehensive numerical study of PINN on solving neutron diffusion eigenvalue problem}
	
	\maketitle

	\begin{abstract}
		We put forward a data-enabled physics-informed neural network (DEPINN) with comprehensive numerical study for solving industrial scale neutron diffusion eigenvalue problems (NDEPs). In order to achieve an engineering acceptable accuracy for complex engineering problems, a very small amount of prior data from physical experiments are suggested to be used, to improve the accuracy and efficiency of training. We design an adaptive optimization procedure with Adam and LBFGS to accelerate the convergence in the training stage. We discuss the effect of different physical parameters, sampling techniques, loss function allocation and the generalization performance of the proposed DEPINN model for solving complex eigenvalue problems. The feasibility of proposed DEPINN model is verified on three typical benchmark problems, from simple geometry to complex geometry, and from mono-energetic equation to two-group equations. Numerous numerical results show that DEPINN can efficiently solve NDEPs with an appropriate optimization procedure. The proposed DEPINN can be generalized for other input parameter settings once its structure been trained. This work confirms the possibility of DEPINN for practical engineering applications in nuclear reactor physics.	
	\end{abstract}


	%
	\IEEEpeerreviewmaketitle


	\section{Introduction}
	\label{sec:intro}
	
	In the field of nuclear reactor engineering, it is essential to predict how the neutrons will be distributed throughout the reactor core. This is also a highly difficult problem because the neutrons interact differently with different materials in a reactor core. The neutron diffusion theory provides a basis for neutron-physical simulation of nuclear cores, and is widely used in practical applications. There are relatively mature numerical calculation methods for solving the neutron diffusion equations, particularly  neutron diffusion eigenvalue problems (NDEPs) in the industry, mainly including finite difference method, different kinds of node methods and so on \citep{Hebert2009}.

	The development of scientific machine learning (SML) provides and offers another way to solve nuclear engineering problems with mesh free, easy implementation properties. 
	\textcolor{black}{Many related work can be found in thermal-hydraulics domain such as boiling heat transfer \cite{LIU2018305}, turbulence model for reactor transient analysis \cite{liuyang2021ASME}, multiphase-CFD simulations of bubbly flows \cite{LIU2021107636} and critical heat flux prediction \cite{ZHAO2020114540} etc. The developments of SML in nuclear reactor physics domain, solving neutron diffusion equations, are relatively few. The work in \cite{Zhangqian2021} used a convolutional neural networks (CNN) to build a deep learning based surrogate model for estimating the flux and power distribution solved by a diffusion equation. In that work, the flux and power distribution calculated by neutron simulation code are used for training, which means the surrogate model is still industrial-code-dependent. The work \cite{jne2040036} went further, and proposed a physicals-informed neural network based deep learning framework to solve multi-dimensional mono-energetic neutron diffusion problems.
	}
	Recently, the work in \citep{PINN-MDNDE} explored the possibility of physics-informed neural networks (PINN) \citep{PINN} in solving NDEPs with numerical test on several one dimensional mono-energetic problems. 
	\textcolor{black}{In this aspect, much more comprehensive numerical study based on complex geometry and two-group diffusion equations - which is widely used in industry - should be brought before the practical application of the learning problem into reactor physics domain.
	}

	
	\subsection{The development of scientific machine learning}
	\label{sec:Scientific}
	In recent years, with the rapid improvement of computing resources, artificial intelligence technology represented by deep learning has made major breakthroughs in image processing, natural language processing. However, deep learning still has no mature application scenarios in the field of computational physics and computational mathematics. Using deep learning to solve partial differential equations (PDEs) is still a frontier research hotspot.
	Traditional numerical methods, such as finite difference method and finite element method, have high dependence on grids and may cause curse of dimensionality \citep{DP-COD}, which can be avoided by deep neural network (DNN) \citep{CofD} and automatic differentiation technology \citep{autodiff}.

	
	Up to now, scholars have proposed many methods for deep learning in computational science, such as the strong form of PDEs (\citep{PINN}, \citep{DGM}), the variational form of PDEs (\citep{DeepRitz}), method for stochastic differential equations (SDEs) \citep{DeepBSDE} and operator learning of PDE coefficients (\citep{DeepONet}, \citep{FNO}). We particularly focus on the PINN \citep{PINN} proposed by M. Raissi et al. In many fields of physics and engineering, some prior information is often implied in the PDE to be solved, such as the solution satisfies additional properties in certain computational regions. PINN can train models with prior information which is not considered in any other DNN. Compared with traditional numerical methods, PINN has the advantages of independent of grids, which can be used to solve high-dimensional problems and inverse problems.  Also, it has strong generalization ability.
	
	
	Once PINN has been proposed, it has got extensive attention from researchers. Scholars have also made a very thorough analysis and improvement on some aspects of PINN. In the aspect of generalization error estimation(\citep{PINNEE1}), they have proved the convergence of the solution obtained by training PINN under certain regularity assumptions. In the aspect of training process (\citep{PINNTA1}, \citep{PINNTA2}, \citep{PINNTA3}), the reasons for the failure of PINN training are analyzed from the aspects of internal weight of loss function, neural tangent kernel, and frequency of solution.

	In addition, researchers have also done a great deal of contributions on how to use PINN to solve eigenvalue problems.  For example, in \citep{PINN_EP1}, forced boundary conditions and exponential loss function are used to enable the neural network to learn non-trivial solutions and appropriate eigenvalue. In \citep{PINN_EP3}, the minimum eigenvalue is searched by adding the Rayleigh quotient to the loss function, which is advantageous to solve for eigenvalues of high-dimensional PDEs. In \citep{PINN-MDNDE}, a parallel search method for eigenvalues is proposed, and the parameter sensitivity of PINN on single-group problems is explored based on simple one dimension problems.

	
	\subsection{Contribution of this work}
	\label{sec:Contribution}
	\textcolor{black}{
		To push further based on the prior work  \citep{PINN-MDNDE, jne2040036}, the feasibility of PINN on two-group diffusion equations problems with complex geometry shall be investigated rather than testing just on analytical problems. There are at least three issues should be solved before the practical application of the learning problem into reactor physics domain:
		\begin{itemize}
			\item \textcolor{black}{How well the loss function is designed for making full use of all possible information to learn the physical model}.
			\item How well the sampling of the data describes the true physical model.
			\item How well the optimization procedure performs in finding the best fits to the sampled data.
		\end{itemize}
	} 
	
	In this work, we bring out a systematic analysis on the overall performance of PINN in solving NDEPs. \textcolor{black}{Furthermore, in order to distinguish ourselves from the common data-driven technology and emphasize the effective use of a small amount of data simulated from experiments, we propose a data-enable physics-informed neural network (DEPINN) to achieve an engineering acceptable accuracy for complex engineering problems.} We train a DEPINN on a variety of datasets stemming from different parameter choices of the eigenvalue problem of neutron diffusion equations,
	\textcolor{black}{based on benchmark problems from simple geometry to complex geometry, and from mono-energetic to two-group equations. Specifically, we test our proposed DEPINN framework in solving the following three problems: i) the finite spherical reactor modeled by one-dimension mono-energetic diffusion equation, ii)  the finite cylindrical reactor modeled by two-dimension mono-energetic diffusion equation, and iii) the 2D IAEA Benchmark Problem (IBP) \citep{Benchmark} modeled by two-dimension two-group diffusion equations, which was adapted from a practical nuclear reactor. 
	}
	
	The experiment is designed such that we vary the relationship between the capacity of the architecture and the complexity of the data and report the effect on the overall performance. In designing such an experiment, we face the aforementioned three fundamental challenges in design a practical neural networks for solving the neutron diffusion equations: (i) effect of the loss function allocation, (ii) effect of the optimization procedure, (iii) effect of the sampling procedure.
	Particularly, \textcolor{black}{we use DEPINN to approximate the solution of NDEPs with a few prior data in case of complex geometry}.  We design an adaptive optimization method in the training processing, sampling techniques and discuss the effect of different physical parameters, which shown that our method is very efficient for the engineering problem.
	\textcolor{black}{In addition, all data and codes used in this manuscript are publicly available on GitHub.
	}
	
	The rest of this article is organized as follows. Section II describes the mathematical modeling of parametric neutron diffusion eigenvalue problem. In Section III, the implementation and adaption of PINN are introduced.  Numerical results with application to several benchmark problems are illustrated in Section IV. Finally, Section V summarizes and concludes this article.

	\section{The parametric neutron diffusion eigenvalue problem}
	\label{sec:The parametric neutron diffusion eigenvalue problem}
	
	In the simulation of nuclear reactor core, the neutron flux $\phi=\left( \phi_1,\phi_2 \right)^{T}$ is usually modeled by two-group neutron diffusion equations with suitable boundary conditions. Index 1 and 2 denote the high and thermal energy group respectively.  The flux is the solution to the following eigenvalue problem (see \cite{Hebert2009}). To be precise, the flux $\phi$ satisfies the following eigenvalue problem:
	Find $\left(\lambda,\phi \right)\in \mathbb C\times L^\infty(\Omega)\times L^\infty(\Omega)$, s.t.
	\begin{small}
		\begin{equation}
			\label{eq:diffusion}
			\hspace{-0.3cm}
			\begin{array}{r@{}l}
				\left\{
				\begin{aligned}
					-\nabla \left(D_1\nabla \phi_1\right)+\left(\Sigma_{{ a}, 1} + \Sigma_{ 1\to 2} \right) \phi_1& =
					\lambda \chi_1 \left( \nu\Sigma_{{ f}, 1}\phi_1+\nu \Sigma_{{ f}, 2}\phi_2 \right) \\
					-\nabla \left(D_2\nabla \phi_2\right)+\Sigma_{{\rm a}, 2}\phi_2 - \Sigma_{ 1\to 2}  \phi_1& =
					\lambda \chi_2 \left( \nu\Sigma_{{ f}, 1}\phi_1+ \nu\Sigma_{{ f}, 2}\phi_2 \right)
				\end{aligned}
				\right.
			\end{array}
	\end{equation}
\end{small}
with
\begin{small}
	\begin{equation}
		\label{eq:PINN_eigenvalue BC}
		\begin{array}{r@{}l}
			\left\{
			\begin{aligned}
				&\frac{\partial \phi_h(\bx)}{\partial \bn} = 0, \ \forall \bx \in \partial \Omega_{L}, \\
				&\frac{\partial \phi_h(\bx)}{\partial \bn} + \alpha \phi_{h} = g(\bx) ,\  \forall \bx \in \partial \Omega_{R}, \\
			\end{aligned}
			\right.
		\end{array}
	\end{equation}
\end{small}
where $h = 1$ or $2$, $\partial \Omega = \partial \Omega_{L} \bigcup \partial \Omega_{R}$, $\alpha$  and $g$ are given parameters. The generated nuclear power is $P=\nu\Sigma_{{ f}, 1}\phi_1+ \nu\Sigma_{{ f}, 2}\phi_2.$
%
%
%
The following parameters are involved in the above equation: $\Sigma_{1\to 2}$ is the macroscopic scattering cross section from group $1$ to $2$; $D_i$ is the diffusion coefficient of group $i$ with $i\in \{1,2\}$;  $\Sigma_{{ a}, i}$, $\Sigma_{{f}, i}$ and $\chi_i$ are the macroscopic absorption cross section, macroscopic fission cross section and the fission spectrum of group $i$ respectively; $\nu$ is the average number of neutrons emitted per fission.
We make some comments on the coefficients and recall well-posedness results of the eigenvalue problem Eq. \eqref{eq:diffusion}. First of all, the first four coefficients ($D_i$, $\Sigma_{a,i}$, $\Sigma_{s,1\to 2}$ and $\Sigma_{f,i}$) might depend on the spatial variable. In the following, we assume that they are either constant or piecewise constant so that our set of parameters is $\mu=\{ D_1, D_2, \Sigma_{a,1}, \Sigma_{a,2}, \Sigma_{1\to2}, \nu, \Sigma_{f,1}, \Sigma_{f,2}, \chi_1, \chi_2 \}.$
%


Under some mild conditions on the parameters $\mu$, the maximum eigenvalue $\lambda_{\text{max}}$ is real and strictly positive (see \citep[Chapter XXI]{DLvol6}). 
The associated eigenfunction $\phi$ is also real and positive at each point $\mathbf x\in\Omega$ and it is the flux of interest. 
In neutronics, it is customary to use the inverse of $\lambda_{\text{max}}$, that is called the multiplication factor $k_{\text{eff}} := \frac{1}{\lambda_{\text{max}}}$.
Here, for each setting $\mu$, the parameter $k_{\text{eff}}$ is determined by the solution to the eigenvalue problem Eq. \eqref{eq:diffusion}. 




\section{Implement of neural networks}
\label{sec:Realizations}

\subsection{Introduction of DEPINN}
\label{sec:PINN Introduction}
In this section, we introduce how to use PINN with prior data, i.e. DEPINN, to solve the problems \eqref{eq:diffusion}-\eqref{eq:PINN_eigenvalue BC}. PINN \citep{PINN} is one of the most well-known methods for solving PDEs using DNN. The main idea is to train DNN as an approximator. The output is an approximate solution to the equation, which is achieved by incorporating such as PDEs, definite solution conditions, related physical laws and prior information into the construction of the loss function of the DNN. Then the neural network is generally trained by optimization algorithms such as Adam \citep{Adam} and LBFGS \citep{LBFGS}. \textcolor{black}{During the training process, the value of the loss function will gradually decrease. And when the value of the loss function drops to a sufficiently small value, it can be considered that the predicted solution will be close enough to the true solution of the equation.}



Eq. \eqref{eq:diffusion} can be rewritten as
\begin{equation}
	\label{eq:PINNPDE}
	\left\{
	\begin{aligned}
		&\mathcal{N}_1[\phi(\bx),\lambda]=0 , \  \forall \bx \in \Omega,\\
		&\mathcal{N}_2[\phi(\bx),\lambda]=0 ,\  \forall \bx \in \Omega,\\
		& \mathcal{B}[\phi(\bx)]=0 , \ \forall \bx \in \partial \Omega,
	\end{aligned}
	\right.
\end{equation}
where $\Omega$ represents the computational domain, 
and $\mathcal{B}[\phi(\bx)] = 0$ represents the Neumann boundary conditions or mixed boundary conditions, which are from real physics. 


\begin{footnotesize}
	\begin{equation}
		\label{eq:PINNloss}
		\left\{
		\begin{aligned}
			&\mathcal{L}(\bx,\lambda^{\text{NN}};W) = \alpha_1 \mathcal{L}_{res}(\bx,\lambda^{\text{NN}};W) +  \alpha_2 \mathcal{L}_{b}(\bx,\lambda^{\text{NN}};W) +  \\
			&\qquad \qquad  \qquad  \qquad  \alpha_3 \mathcal{L}_p((\bx,\lambda^{\text{NN}};W)),\\
			&\mathcal{L}_{res}(\bx,\lambda^{\text{NN}};W) = \sum_{i} [\mathcal{N}_1[\phi(\bx_i;W),\lambda^{\text{NN}}]^2+\mathcal{N}_2[\phi(\bx_i;W),\lambda^{\text{NN}}]^2] ,\\
			&\mathcal{L}_{b}(\bx,\lambda^{\text{NN}};W) = \sum_{j} (\mathcal{B}[\phi(\bx_j;W)])^2,\\
			&\mathcal{L}_p(\bx,\lambda^{\text{NN}};W) = \sum_{k} [\phi(\bx_k;W) -\phi^{p}(\bx_k)]^2.
		\end{aligned}
		\right.
	\end{equation}
\end{footnotesize}

Now we consider how to construct a DEPINN to solve Eq. \eqref{eq:PINNPDE}.
Let the output of the DNN be $\phi(\bx;W)$, which is the approximate solution of Eq. \eqref{eq:diffusion} and $W$ represents the parameters in the neural networks. Note that $\lambda$ is not the output of the network, we do not have to design a search algorithm for $\lambda$, but treat it as a trainable variable inside the neural networks. We choose fully connected neural networks (FCNN) and use automatic differentiation for all differentiation operations in neural networks. Therefore, the loss function can be described as in Eq. \eqref{eq:PINNloss}.
Where $\lambda^{\text{NN}}$ is the trainable variable corresponding to $\lambda$, $\mathcal{L}_{res}(\bx,\lambda;W)$ is the sum square error (SSE) loss corresponds to the residual of PDE equation, 
$\mathcal{L}_{b}(\bx,\lambda;W)$ is the SSE loss corresponds to the boundary conditions, 
$\mathcal{L}_p(\bx,\lambda;W)$ is the SSE loss corresponds to the prior physical information, \textcolor{black}{$\phi^{\text{P}}(\bx_k)$ represents the prior information obtained from observation or from a finite element solver}, 
$\alpha_1$,$\alpha_2$ and $\alpha_3$ represent the weight of $\mathcal{L}_{res}$, $\mathcal{L}_{b}$ and $\mathcal{L}_{p}$ respectively in the total loss $\mathcal{L}(\bx,\lambda;W)$.

\begin{figure}[htp]
\includegraphics[width=0.55\textwidth]{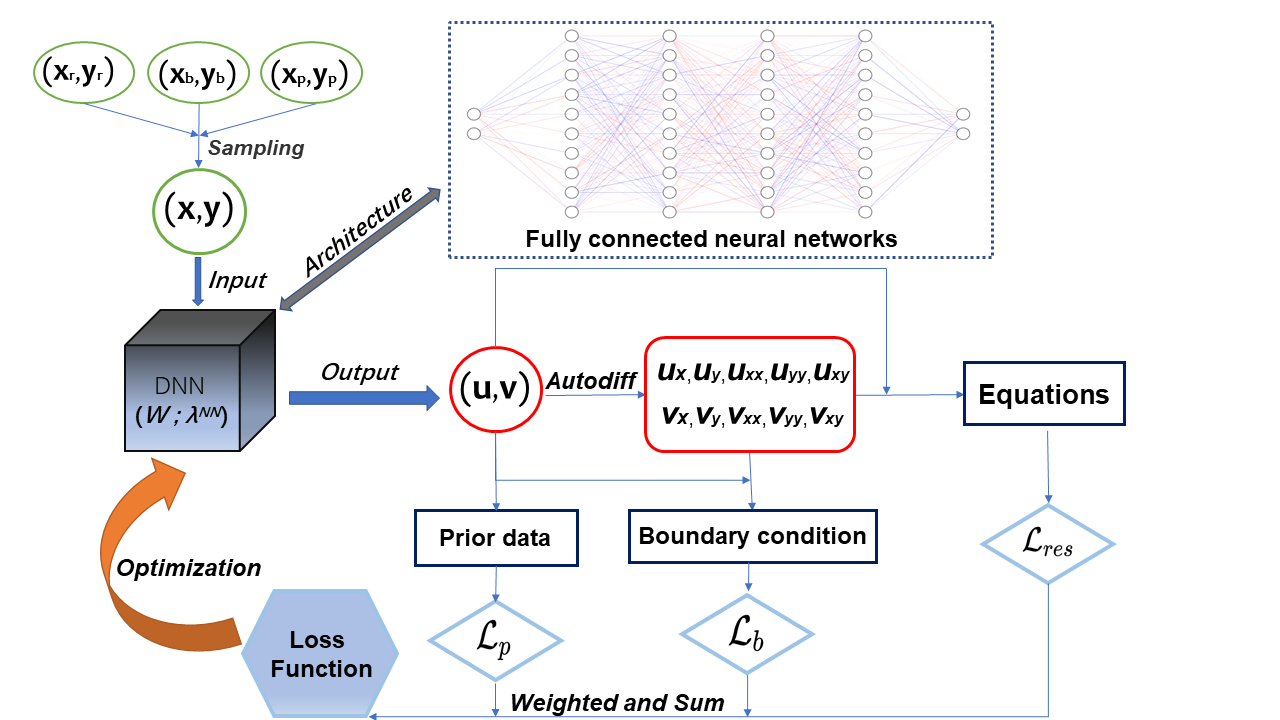}
\centering
\caption{The process of solving the 2D IBP by PINN. }
\label{fig:PINN_Benchmark} 
\end{figure}

Note that our loss function here uses SSE instead of the usual mean squared error (MSE) loss, and the reason is described in the section \ref{sec:lossfunction}, \textcolor{black}{where we conducted rigorous contrast experiments to show} how the weights $\alpha_1$,$\alpha_2$ and $\alpha_3$ affect the results. The value of $\mathcal{L}(\bx,\lambda;W)$ drops to a very small value, meanwhile the parameter $W$ in the neural network is well optimized, which means that $\phi(\bx;W)$ will approximate the real solution sufficiently. The whole process above is illustrated in Fig. \ref{fig:PINN_Benchmark}. \textcolor{black}{Note also that, if no prior data are available, the DEPINN degenerates to general PINN,  which is only constrained by physical information.}

\subsection{Methodology in DEPINN}
\label{sec:PINN Methodology}

\subsubsection{Few shot data-driven}
\label{sec:few data driven}

\textcolor{black}{There are many ways to obtain the prior information, for example, it can be extracted from the analytic solution, obtained through physical constraints or prior inference, or obtained through the detection of detectors in real experiments (e.g. flux or power measured in the core).} Assume we only have a very small number of prior points $(x_p,y_p)$ and prior solutions $(\phi^{\text{P}}_1(x_p,y_p), \phi^{\text{P}}_2(x_p,y_p))$ for data-driven, which are called prior data. These data are crucial for DEPINN to solve complex neutron diffusion problem, and it has three main functions: (i) A unique solution is obtained from the network with prior data. 
(ii) 
We do not need to design a search algorithm for $k_{\text{eff}}$, since prior data not only shows the value of $\phi_1$ and $\phi_2$ at a certain point $(x_p, y_p)$, but also contains the information of $k_{\text{eff}}$.
(iii) Prior data-driven can accelerate the optimization process of DEPINN. After initialization, the gradient of the loss term $\mathcal{L}_p(\bx,\lambda;W)$ in the loss function is the largest, but descends the fastest (Tab. \ref{tab:DLT_Benchmark}). \textcolor{black}{Through \ref{tab:PredictComparison}, it can be seen intuitively that the prediction performance of the network has been greatly improved in each training period due to the addition of prior data. Furthermore, according to the phenomena shown in Table 1,2 and the results we measured in these two experiments, we found that with the rapid decline of the prior loss term at the beginning of the training, the prediction error at prior points also dropped faster than the ordinary residual point. Therefore, We think that the addition of prior data makes the original optimization problem become a conditional extremum problem constrained by prior data. Note also that, at the end of the training process, the boundary condition loss term dominates the total loss term. This phenomenon reflects the network is sensitive to the boundary conditions, which provide a direction for the further improvement of the proposed DEPINN. In Fig. \ref{fig:IAEA_Heatmap}, we can find that  relative large error appears around the interface, which also confirms this phenomenon.}



\begin{table}[htp] 
\footnotesize
\caption{
	The value(proportion) of different loss terms with different training epochs (solving 2DIBP).
}
\begin{tabular}{c|ccc}
	\hline\noalign{\smallskip}
	Epochs & $\mathcal{L}_{res}(\bx,\lambda;W)$ & $\mathcal{L}_{b}(\bx,\lambda;W)$ & $\mathcal{L}_{p}(\bx,\lambda;W)$ \\
	\hline
	$10$ &  239.28($\boldsymbol{0.47\%}$) &   374.29($\boldsymbol{0.73\%}$)  & 50323.74($\boldsymbol{98.80\%}$)  \\
	\hline
	$2000$ &  1412.34($\boldsymbol{43.79\%}$) &  1369.20($\boldsymbol{42.45\%}$)  & 443.74($\boldsymbol{13.76\%}$)  \\
	\hline
	$5000$ & 4.22($\boldsymbol{4.52\%}$) & 87.77($\boldsymbol{94.00\%}$) & 1.38($\boldsymbol{1.48\%}$) \\
	\hline
	$15000$ &  1.15($\boldsymbol{3.23\%}$) & 34.37 ($\boldsymbol{96.64\%}$) & 0.05($\boldsymbol{0.13\%}$) \\
	\hline
	$30000$ & 1.01($\boldsymbol{3.63\%}$) & 26.71 ($\boldsymbol{96.21\%}$) & 0.05($\boldsymbol{0.17\%}$)  \\
	\hline
\end{tabular}
\label{tab:DLT_Benchmark}
\end{table}
\begin{table}[h]
\centering
\caption{
	\textcolor{black}{Comparison of prediction performance of neural networks with or without prior information at different training epochs (solving 2DIBP). The upper layer has no prior information, while the lower layer has. The measurements in the table can be found in Section IV.B. Without prior information, the optimization algorithm stopped working after $\sim$50000 epochs. Therefore, the upper layer data were not displayed at 80000 epochs.}
}
\setlength{\tabcolsep}{1.0mm}{
	\begin{tabular}{c|c|c|c|c|c}
		\hline\noalign{\smallskip}
		Epochs& $e_{\infty}(u)$  & $e_{\infty}(v)$ & $e_{2}(u)$ & $e_{2}(v)$ &   $e(k^{\text{NN}}_{\text{eff}})$\\
		\hline
		10 & 0.9478 & 1.0000 & 0.9522&	0.9881& 8.9439\\
		&  $\boldsymbol{0.8872}$ & $\boldsymbol{1.0035}$ & $\boldsymbol{0.8959}$&	$\boldsymbol{0.9671}$& $\boldsymbol{8.5499}$\\
		\hline
		10000&  0.7137& 1.0003	&0.8034 & 0.8691& 0.02715\\
		&  $\boldsymbol{0.0668}$&	$\boldsymbol{0.3487}$	& $\boldsymbol{0.0191}$& $\boldsymbol{0.1038}$ & $\boldsymbol{0.0021}$\\
		\hline
		30000&  0.6423 & 1.0001	&0.6813	&0.7903&0.0137 \\
		&  $\boldsymbol{0.0495}$ & $\boldsymbol{0.2670}$	&$\boldsymbol{0.0137}$	&$\boldsymbol{0.0885}$&$\boldsymbol{0.0016}$\\
		\hline
		50000 & 0.6290&	1.0001&	0.6575&	0.7751& 0.0118\\
		&$\boldsymbol{0.0396}$& $\boldsymbol{0.1260}$ & $\boldsymbol{0.0115}$& $\boldsymbol{0.0541}$& $\boldsymbol{0.0001}$\\
		\hline
		80000&	-  &- &	 -&	- & - \\
		&	$\boldsymbol{0.0110}$&	$\boldsymbol{0.0331}$&	$\boldsymbol{0.0029}$&	$\boldsymbol{ 0.0106}$&$\boldsymbol{0.0002}$\\
		\hline
	\end{tabular}
}
\label{tab:PredictComparison} 
\end{table}

\subsubsection{Adaptive optimization method combining Adam and LBFGS}
\label{sec:Adaptive training method}

\begin{figure}[htbp]
\centering
\subfloat[Total Loss $\mathcal{L}$ during training ]{\includegraphics[width = 0.24\textwidth]{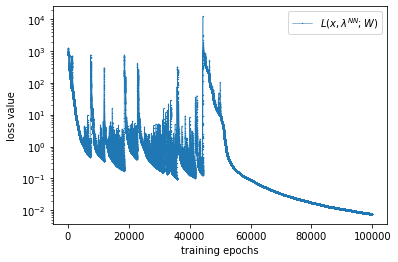}}
\subfloat[Total Loss $\mathcal{L}$ during training]{\includegraphics[width = 0.24\textwidth]{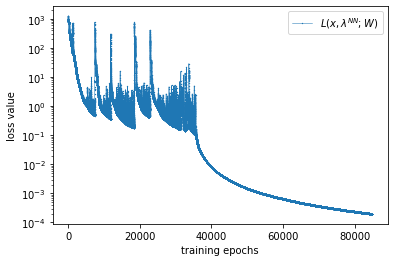}}\\
\subfloat[ Relative  $L_{\infty}$ error of $u$ and $v$]{\includegraphics[width = 0.24\textwidth]{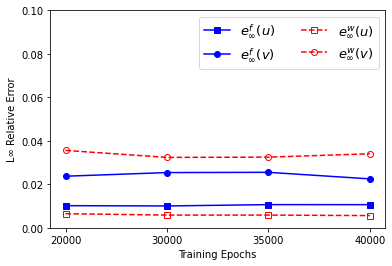}}
\subfloat[Relative error of $ k^{\text{NN}}_\text{eff}$]{\includegraphics[width = 0.24\textwidth]{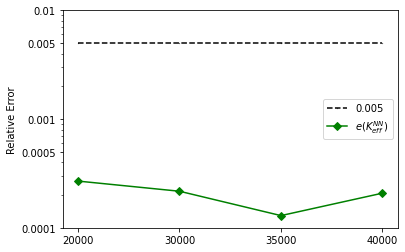}}
\caption{
	(a) Train Adam algorithm 50000 epochs and LBFGS algorithm 50000  epochs, (b) Train Adam algorithm 35000 epochs and LBFGS algorithm 50000  epochs. Due to the randomness of Adam algorithm, if LBFGS algorithm is used when the loss function fluctuates to a large value, the result will be worse. (c)-(d) show the variations in the relative error of the output flux $\phi(x,y;W)$ and $ k^{\text{NN}}_\text{eff}$ in the 2D IBP when the Adam training epochs are fixed at 25,000, 30,000, 35,000, and 40,000, then using the LBFGS algorithm for 50,000 times.} 
\label{fig:AdamProblem}
\end{figure}

Generally, two optimization methods, i.e., Adam and LBFGS, are used to optimize DEPINN \cite{PINN}, in which Adam algorithm is used with $T_1$ steps, then LBFGS algorithm is used until the stop criteria is achieved. However, fixing $T_1$ epochs for Adam algorithm may cause some problems: (i) if $T_1$ is too small, Adam algorithm may not fully play the role of optimization, and the value of the loss function passed to the LBFGS algorithm will be large, which will require more training epochs to achieve the stop criteria. If $T_1$ is too large, the optimization effect in the later stage of Adam algorithm is not significant (Fig. \ref{fig:AdamProblem}(c)-(d)), due to that Adam algorithm is a stochastic gradient descent algorithm. A timely connection to LBFGS algorithm is necessary. (ii) 
If the training of Adam algorithm is ended when the value of the loss function fluctuates to a relatively large value, the result is very bad (Fig. \ref{fig:AdamProblem}(a)-(b)). 

Therefore, in order to overcome the above difficulties, it is necessary to design an adaptive algorithm to switch Adam algorithm to  LBFGS algorithm automatically. The adaptive algorithm is presented, which ensures the loss function does not fluctuate drastically in the last $S_1$ training epochs, also ensures that the decline of the loss function is not obvious, therefore the training is finished efficiently. The optimization procedure we proposed is described as following. 

%
%

We first introduce two symbols $S_1$ and $\epsilon_1$. For a loss function that has been optimized for $\bt$ times training by Adam, if $\bt$ is an integer multiple of $S_1$ and is large enough, then take the difference of the value of the loss function of the last 2$S_1$ training and then take the average value of the absolute value of the difference result every $S_1$ times, we can get two difference averages $\overline{\text{dif}}_1$,$\overline{\text{dif}}_2$. If the difference of the two absolute values $\overline{\text{dif}}_1 - \overline{\text{dif}}_2$ drops below $\epsilon_1$, Adam training ends and LBFGS algorithm starts working. After obtaining an appropriate threshold $\epsilon_1$, by making the absolute value of the two difference averages less than $\epsilon_1$, not only can it ensure that the loss function does not fluctuate drastically in the last $S_1$ training epochs, but it can also ensure that the decline of the loss function is not obvious, which shows that Adam optimization has also fully worked. The above process can be seen in Alg. \ref{algo:1}.

\vskip -0.3cm
\begin{algorithm}[!htbp]
\caption{Adaptive optimization method with Adam and LBFGS for DEPINN}
\text{Symbols}: \small{$\bt$ means the current training epochs, $T_1$ means maximum number of training epochs, \\
$S_1$ and $\epsilon_1$ are two hyperparameters in this algorithm,\\
$\mathcal{L}^{\bt}$ is a t-dimensional vector representing the value of the loss function in the previous $\bt$ training times, \\
$\text{A}[m:n]$ means from the m-th component to the n-th component in the vector $\text{A}$, $\text{mean}(\text{A})$ means take the average of the vector $\text{A}$.\\}
\While{$\bt<T_1$}{
Run the t-th Adam optimization algorithm \\
\eIf{$S_1 | \bt$ \text{and} $\bt>2S_1$}
{Let $i=\bt-2S_1-1$\\
	\While{$i< \bt $}{
		$\text{dif}(i) = \lvert \mathcal{L}^{\bt}(i) - \mathcal{L}^{\bt}(i+1) \lvert$  \\
		$i=i+1 $
	}
	$\overline{\text{dif}}_1 =\text{mean}(\text{dif}[\bt-S_1:\bt]) $ \\
	$\overline{\text{dif}}_2 =\text{mean}(\text{dif}[\bt-2S_1:\bt-S_1]) $ \\
	\If{$\lvert \overline{\text{dif}}_1 - \overline{\text{dif}}_2 \rvert < \epsilon_1$}
	{Break\\
		Start LBFGS optimization algorithm}
}
{$t=t+1$}
}
\label{algo:1}
\end{algorithm}

\section{Numerical survey on parameter settings of NEPINN}
\label{sec:Numericalsurvey}
\begin{figure}[htp]
\includegraphics[width=0.45\textwidth]{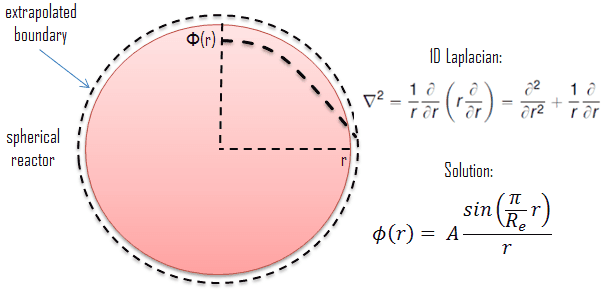}
\centering
\caption{Geometry of finite spherical reactor \cite{benchmark1D}.}
\label{fig:1dspherical} 
\end{figure}

\begin{figure}[htp]
\includegraphics[width=0.45\textwidth]{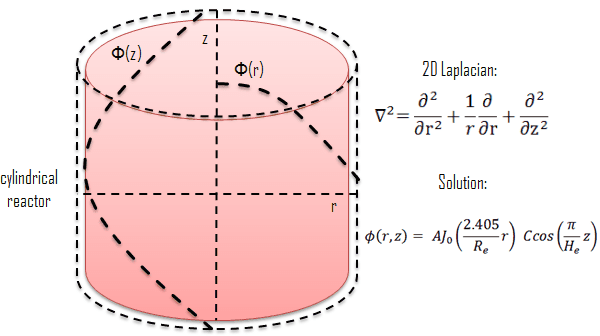}
\centering
\caption{Geometry of finite cylindrical  reactor \cite{benchmark1D}.}
\label{fig:2dcylindrical} 
\end{figure}

\begin{figure}[htp]
\includegraphics[width=0.35\textwidth]{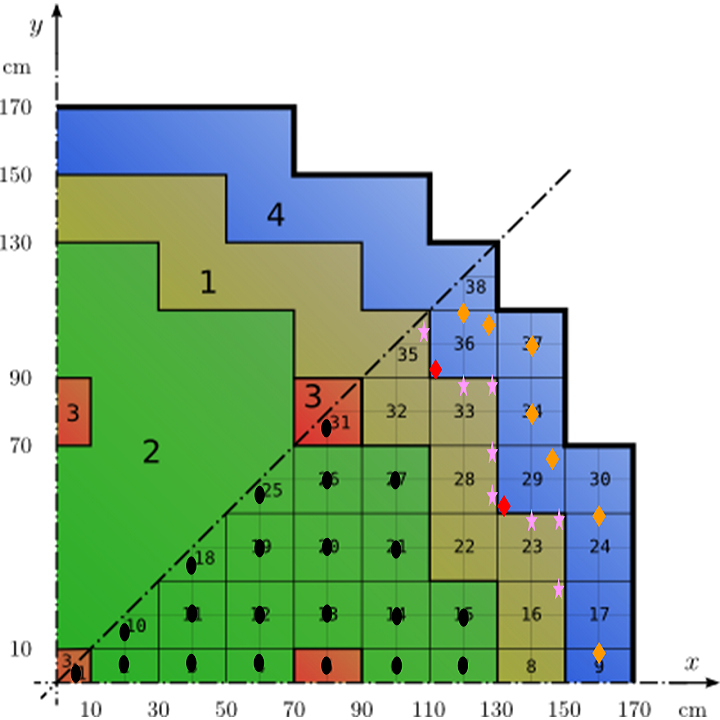}
\centering
\caption{
Geometry of 2D IBP, upper octant: region assignments, lower octant: fuel assembly identification \cite{2D-IAEA-Benchmark}.
}
\label{fig:iaeacore} 
\end{figure}

\textcolor{black}{
In this section, we will illustrate the efficiency of the proposed DEPINN framework for solving NDEPs based on three benchmark problems. We give an detail introduction of these benchmark problems in section \ref{sec:Setup}, and we give the evaluation measures of the article and the engineering acceptance criteria in section \ref{sec:measures}. Considering the importance of hyperparameters in DEPINN, we explore in detail the impact of the loss function configuration (Section \ref{sec:lossfunction}), optimization strategy (Section \ref{sec:optimization}) and sampling strategy(Section \ref{sec:sampling}) on DEPINN. Finally, we explore the generalization performance of our networks in Section \ref{sec:parameter}.
}
\textcolor{black}{
\subsection{Setup of experiments for benchmark problems}
\label{sec:Setup}
In this subsection, we give an detail introduction of the benchmark problems that will be used in this work, from simple geometry to complex geometry, and from mono-energetic equation to two-group equations.
\subsubsection{Case-1: the finite spherical reactor.}
This test case is adapted from \cite{benchmark1D} and modeled by one-dimension mono-energetic diffusion equation as shown in Eq. \eqref{eq:1dspherical}. This is a multiplying system of a uniform reactor in the shape of a sphere of physical radius $R$. 
To solve the diffusion equation, we replace the Laplacian with its spherical form as shown in Fig. \ref{fig:1dspherical} with zero flux boundary condition at the extrapolated length, where $R_{e}=R+d$, and $d \approx \frac{2}{3}\lambda_{tr}$ is known as the extrapolated length. 
In this test case, the analytical solution of Eq. \eqref{eq:1dspherical} can be represented as in Eq. \eqref{eq:1dsphericalsolution} when $k_{\text{eff}}=1$. For numerical solution convenience, we further set $\Sigma_a=0.45, \Sigma_s = 2, \Sigma_f=2$ and $\nu\Sigma_f=2.5$.
\begin{equation}
\label{eq:1dspherical}
\hspace{-0.3cm}
\begin{array}{r@{}l}
	\left\{
	\begin{aligned}
		& D \Delta \phi(r) - \Sigma_{a}\phi(r) =\frac{1}{k_{\text{eff}}} \nu \Sigma_f\phi(r) \\
		& \phi(R_{e})= 0,~ r \in (0,R_{e})
	\end{aligned}
	\right.
\end{array}
\end{equation}
}
\begin{equation}
\label{eq:1dsphericalsolution}
\phi(r)=A\frac{\sin(\frac{\pi}{R_{e}}r)}{r}. 
\end{equation}
\textcolor{black}{
\subsubsection{Case-2: the finite cylindrical reactor.}
This test case is also adapted from \cite{benchmark1D}, a multiplying system of a uniform reactor in the shape of a cylinder of physical radius $R$ and height $H$.  The neutronic behaviour can modeled by two-dimension mono-energetic diffusion equation as shown in Eq. \eqref{eq:2dcylindrical}, where $R_{e}=R+d$, $H_{e}=H+d$ and $d$ is the extrapolated length. The parameters are kept the same with test case-1. 
To solve the diffusion equation, we replace the Laplacian by its cylindrical form as shown in Fig \ref{fig:2dcylindrical}. In this case, the analytical solution of Eq. \eqref{eq:1dspherical} can be represented by Eq.\eqref{eq:2dcylindricalsolution} when $k_{\text{eff}}=1$, where $J_0(r)$ is the zero order first kind Bessel function. 
\begin{equation}
\label{eq:2dcylindrical}
\hspace{-0.3cm}
\begin{array}{r@{}l}
\left\{
\begin{aligned}
	& D \Delta \phi(r,z) - \Sigma_{a}\phi(r,z) =\frac{1}{k_{\text{eff}}} \nu \Sigma_f\phi(r,z) \\
	& \phi(R_{e})= 0,~ \phi(H_{e})= 0, ~ r \in (0,R_{e}), z \in (0,H_{e})
\end{aligned}
\right.
\end{array}
\end{equation}
\begin{equation}
\label{eq:2dcylindricalsolution}
\phi(r)=AJ_0(\frac{2.045}{R_e}r)\cos(\frac{\pi}{H_e}z) 
\end{equation}
}
\textcolor{black}{
\subsubsection{Case-3: the 2D IAEA Benchmark Problem.}
We consider the classical 2D IAEA Benchmark Problem (2D IBP) (p.437 of \citep{Benchmark} modeled by two-dimension two-group diffusion equations (Eq. \eqref{eq:diffusion}), see \citep{Benchmark} for its definition and \citep{2D-IAEA-Benchmark} for implementation with neutronic code. The reactor geometry is shown in Figure \ref{fig:iaeacore}. Only one quarter is given because the rest can be inferred by symmetry along the $x$ and $y$ axis. This one quarter is denoted by $\Omega$ and it is composed of four sub-regions of different physical properties:  Neumann boundary conditions are enforced on the left and the bottom boundaries, and the mixed boundary condition is enforced on the external border (the step boundary). The physics coefficients in Eq. \eqref{eq:diffusion} are listed in Tab. \ref{tab:coefs}. 
}
%
\begin{table*}[h]
\centering
\caption{Parameter values of the IAEA 2D benchmark problem.}
\label{tab:coefs}
\begin{tabular}{ccccccccccl}
\hline 
Region & $D_1 $ & $D_2 $ & $\Sigma_{1\to 2}$ & $\Sigma_{a,1}$ & $\Sigma_{a,2}$  & $\nu \Sigma_{f,1}$ & $\nu \Sigma_{f,2}$ & $\chi_1$ & $\chi_2$& Material \\
& (cm) & (cm) & ($\text{cm}^{-1}$) & ($\text{cm}^{-1}$) & ($\text{cm}^{-1}$) & ($\text{cm}^{-1}$) & ($\text{cm}^{-1}$) & & & \\
\hline
$\Omega_1$ & 1.50 & 0.40 & 0.02 & 0.01 & 0.080 &0.00 & 0.135 & 1 & 0 & Fuel 1  \\
$\Omega_2$ & 1.50 & 0.40 & 0.02 & 0.01 & 0.085 &0.00 & 0.135 & 1 & 0& Fuel 2 \\
$\Omega_3$ & 1.50 & 0.40 & 0.02 & 0.01 & 0.130 &0.00 & 0.135 & 1 & 0& Fuel 2 + Rod \\
$\Omega_4$ & 2.00 & 0.30 & 0.04 & 0.00 & 0.010 &0.00 & 0.000 & 0 & 0 & Reflector  \\
\hline
\end{tabular}
\end{table*}

In this article, the reference solution of the two-dimensional two-group diffusion equations are solved by employing the generic high quality finite elements solver FreeFem++ \citep{freefem} that offers a suitable frame for this kind of problems. The eigenvalue problem (Eq.  \eqref{eq:diffusion}) is solved with the tool in arpack++, the object-oriented version of ARPACK eigenvalue package \citep{lehoucq1998arpack}. The function EigenValue computes the generalized eigenvalue of $Au=\lambda Bu$. The  shifted-inverse method is used by default, with sigma $=\sigma$ the shift of the method. The spatial approximation uses $P_1$ finite elements with a grid of size $h = 1$ {\rm cm}. \textcolor{black}{After obtaining the reference solution, we extract a small number of points from the reference solution as the prior information in order to simulate obtaining the prior information through a few detectors.}
\subsection{Evaluation measures}
\label{sec:measures}
In this section, we discuss how to measure the output of DEPINN and the acceptance criteria widely used in engineering. Using $\lambda^{\text{NN}}$ to represent the trainable variables $\lambda$ in the neural networks and $k^{\text{NN}}_{\text{eff}} := \frac{1}{\lambda^{\text{NN}}}$.  Moreover, we use $u,v$ to denote the output flux of DEPINN, use $\phi^{\text{FF}}_1$, $\phi^{\text{FF}}_2$ and $k^{\text{FF}}_{\text{eff}}$ to denote the high-fidelity numerical solutions, such as finite element solutions.


The classical 2D IBP has two regions which are fuel region $\Omega_{\text{fuel}}$ composed $\Omega_{1,2,3}$ and water region $\Omega_{\text{water}}$ composed only of $\Omega_{4}$, which are shown in Fig. \ref{fig:iaeacore}.
Since relative error in $L_{2}$ norm is always satisfactory and relative error in $L_{\infty}$ norm is more important in engineering,  we present the numerical results in $L_{\infty}$ norm in the following sections. The definition of $L_{\infty}$ norm is given in Eq. \eqref{eq:RE_fueL_infty}. 

\begin{equation}\label{eq:RE_fueL_infty}
\begin{aligned}
&     \be_{\infty}^{\text{f}(\text{w})}(u(\hat{x},\hat{y})) = \frac{\Vert u(\hat{x},\hat{y})-\phi^{\text{FF}}_{i}(\hat{x},\hat{y})\Vert_\infty}{\Vert u(x,y) \Vert_\infty}, \ i = 1, 2, \\
&  \forall (\hat{x},\hat{y})  \in \Omega_{\text{fuel}}\left(\Omega_{\text{water}}\right),  \  \forall (x,y)  \in \Omega,\\
&    \be(k^{\text{NN}}_{\text{eff}}) = \frac{\vert \frac{1}{\lambda^{\text{NN}}}-k^\text{FF}_{\text{eff}}\vert}{\vert k^\text{FF}_{\text{eff}} \vert},
\end{aligned}
\end{equation}
where $\be_{\infty}^{\text{f}(\text{w})}(u)$ is the relative $L_\infty$ error of $u$ in $\Omega_{\text{fuel}}$ or $\Omega_{\text{water}}$, $\be(k^\text{NN}_{\text{eff}})$ is the relative error of $k^{\text{NN}}_{\text{eff}}$.

We need to mention the engineering acceptance criteria for 2D IBP. In fuel assemblies with relative flux higher than 0.9, the flux calculation error is less than $5\%$; in fuel assemblies with relative flux less than 0.9, the flux calculation error is less than $8\%$. The relative error of $k^{\text{NN}}_{\text{eff}}$ is less than 0.005.

\subsection{The architecture of the neural network}
Before starting the experiments below, we also need to determine the architecture of the neural network. According to Section \ref{sec:PINN Introduction},  we use FCNN to build DEPINN, it is necessary to explore the sensitivity of DEPINN to the width and depth of the neural networks.

\begin{figure}[htbp]
\centering
\subfloat[ Relative $L_{\infty}$ error of $u$ and $v$]{\includegraphics[width = 0.23\textwidth]{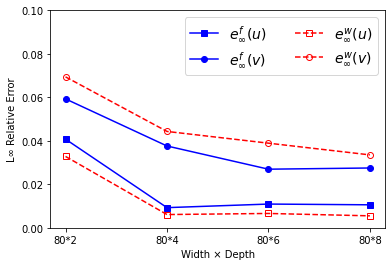}}
\subfloat[Relative error of $k_{\text{eff}}$]{\includegraphics[width = 0.23\textwidth]{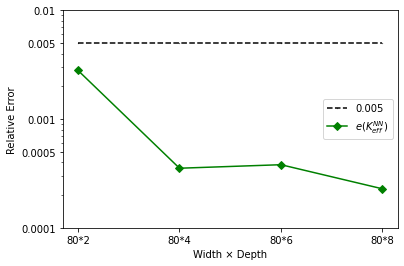}}
\caption{Variations of relative errors of $u$, $v$ and $k_{\text{eff}}$ when net size (width $\times$ depth) takes $80 \times 2$, $80 \times 4$, $80 \times 6$, $80 \times 8$ respectively. (a) The relative error of $u$ and $v$ in $L_\infty$ norm; (b) The relative error of $k_{\text{eff}}$.}
\label{fig:fcnz}
\end{figure}

\begin{table}[h]
\centering
\caption{Other parameter settings in DEPINN.}
\label{tab:parameter-PINN} 
\setlength{\tabcolsep}{1.mm}{
\begin{tabular}{cccccc}
\hline\noalign{\smallskip}
Torch & Cuda & Activation   &  Initialization  & Adam  & LBFGS   \\
version & version & function  & method   &learning rate & learning rate \\
\hline
1.11.0&  10.2& tanh  & Xavier & 0.0001 & 1 \\ 
\hline
\end{tabular}
}
\end{table}

From Fig. \ref{fig:fcnz},  as the depth of the neural network increases, the results gradually become better, which is reasonable. Considering the complexity and generalization ability of the model, subsequent experiments will be carried out with width $80$ and depth $8$. 
\textcolor{black}{Note here that, in this work, we set the depth of the network and the width of each layer experimentally. Further investigation would be needed to explain the physical interoperability of the network.
}
Note also that, all the computations are carried on NVIDIA A100(80G) and NVIDIA TITAN RTX. Other important hyperparameters in PINN are shown in Tab. \ref{tab:parameter-PINN}.

\textcolor{black}{
\subsection{The results of mono-energetic benchmarks}
}
\textcolor{black}{Since Case-1 and Case-2 are much simpler than Case-3, good predictions can be achieved without using data-driven and adaptive optimization methods in Section \ref{sec:PINN Methodology}, thus, the general PINN is used, which is only constrained by physical information. According to Eq. \eqref{eq:1dspherical} and Eq. \eqref{eq:2dcylindrical}, the residual points and boundary points can be sampled randomly from a uniform distribution, and A loss function of the form Eq. \eqref{eq:Case1and2loss} containing only residual items and boundary conditions can be constructed, which will be optimized by Adam and LBFGS algorithms.}

\begin{equation}
\label{eq:Case1and2loss}
\begin{aligned}
\mathcal{L}(\bx;W) = \alpha_1 \sum_{i} [\mathcal{N}[\phi(\bx_i;W)]^2 +  \alpha_2 \sum_{j} (\mathcal{B}[\phi(\bx_j;W)])^2 
\end{aligned}
\end{equation}

\begin{figure}[htbp]
\centering
\subfloat[ Results of Case-1]{\includegraphics[width = 0.23\textwidth]{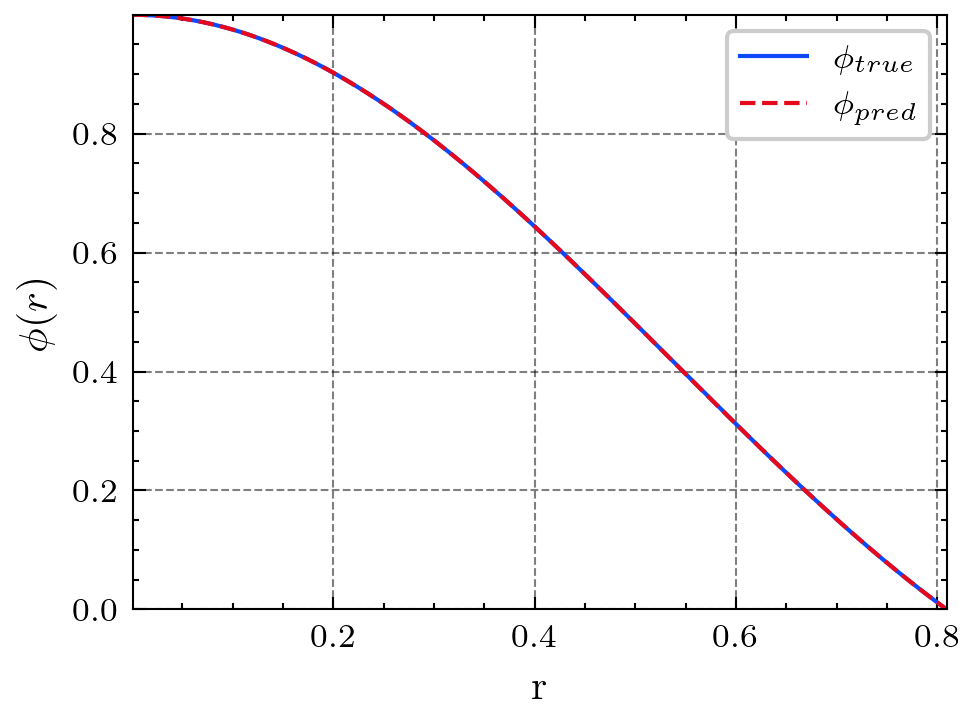}}
\subfloat[ Analytical solution of Case-2]{\includegraphics[width = 0.23\textwidth]{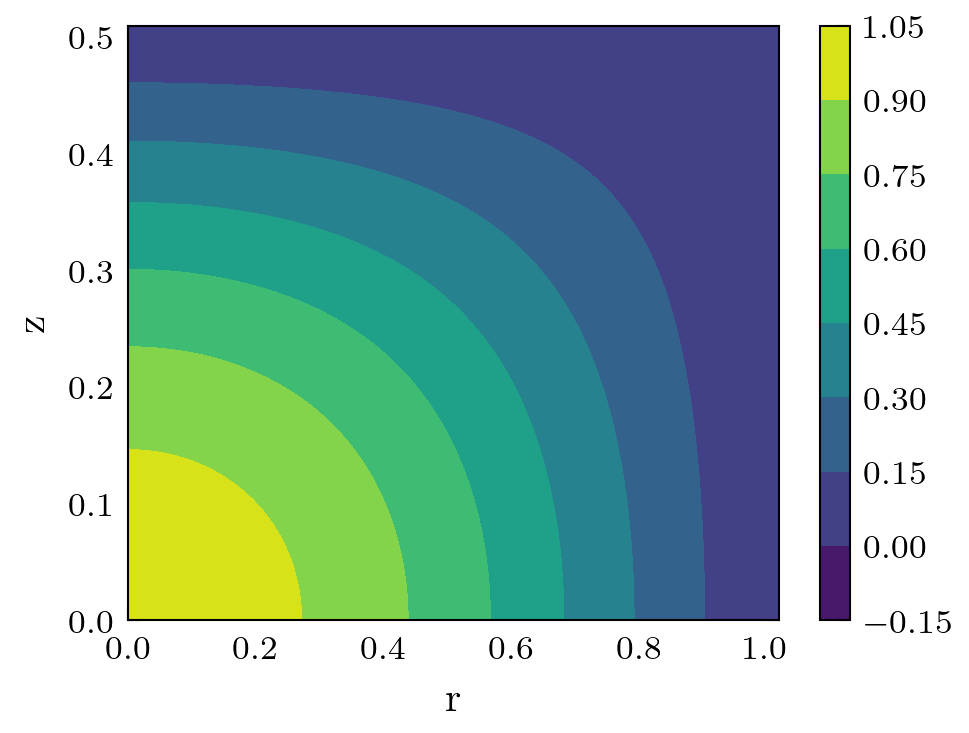}}\\
\subfloat[ Predicted solution of Case-2]{\includegraphics[width = 0.23\textwidth]{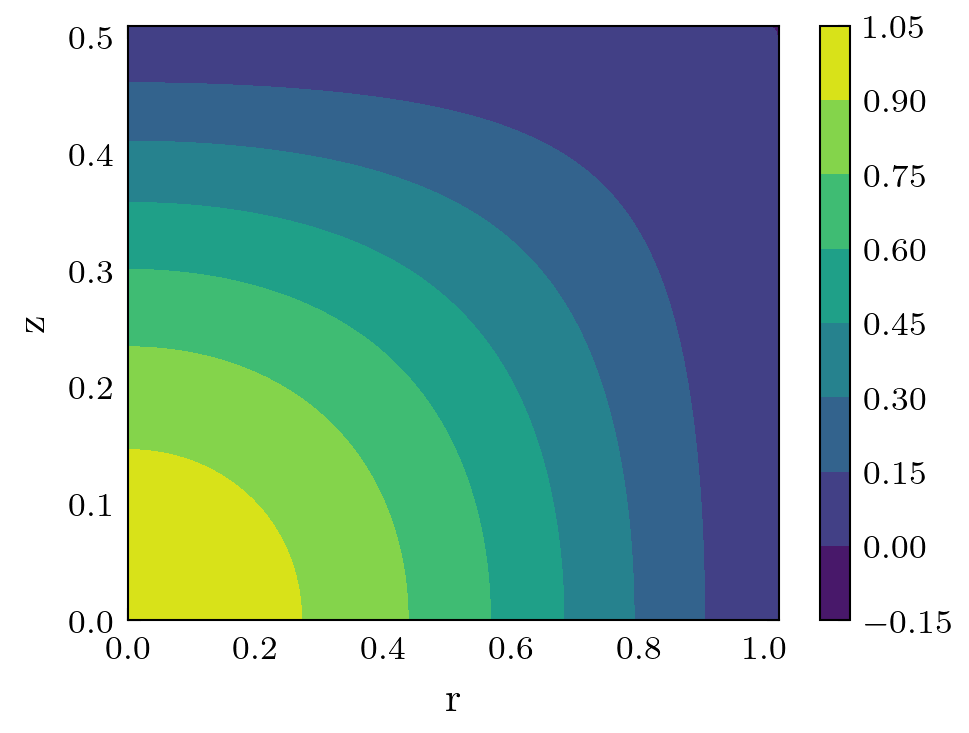}}
\subfloat[ Absolute error of Case-2]{\includegraphics[width = 0.23\textwidth]{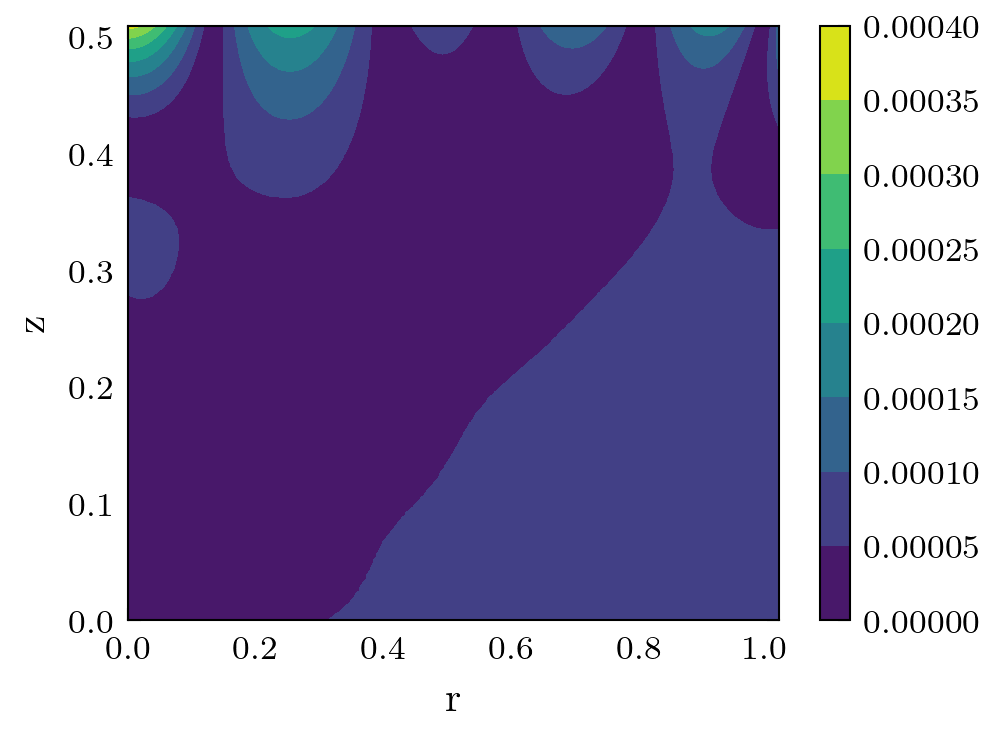}}\\
\caption{(a) shows the predicted output of the neural network in Case-1 and the analytical solution in Eq. \eqref{eq:1dspherical}. The solid blue line is the analytical solution and the dashed red line is the predicted solution. (b)(c)(d) show the results of Case-2 with three heat-maps. (b) shows the analytical solution in Eq. \eqref{eq:2dcylindrical}, (c) shows the predicted solution of the neural network, (d) shows the absolute error between the analytical solution $\phi_{a}$ and the predicted solution $\phi_{p}$, calculated as $|\phi_{p} - \phi_{a}|$.}
\label{fig:Case1and2}
\end{figure}

\begin{table}[h]
\centering
\caption{
Results of mono-energetic benchmarks.
}
\setlength{\tabcolsep}{1.mm}{
\begin{tabular}{c|c|c|c|c}
\hline\noalign{\smallskip}
& $e_{\infty}(\phi_p)$   &  $e_{2}(\phi_p) $ & $\text{MSE}(\phi_p)$   & $\text{APE}(\phi_p)$ \\
\hline
Case-1 & 3.7615e-05 & 1.7018e-05 & 1.3060e-10 & 7.8186e-06 \\
\hline
Case-2  & 3.6901e-04 & 1.2004e-04  & 3.4121e-09 & 4.7293e-05 \\
\hline
\end{tabular}
}
\label{tab:Case1and2_Results} 
\end{table}

\textcolor{black}{Four measurements are given in Tab. \ref{tab:Case1and2_Results}. $e_{\infty}$ and $e_{2}$ represent the relative error under $L_{\infty}$ norm and $L_2$ norm which are defined in Eq. \eqref{eq:Case1and2_Error}: 
\begin{equation}\label{eq:Case1and2_Error}
\begin{aligned}
\be_{\infty}(\phi_p) = \frac{\Vert \phi_p -\phi_a \Vert_\infty}{\Vert\phi_a  \Vert_\infty},   \be_{2}(\phi_p) = \frac{\Vert \phi_p -\phi_a \Vert_2}{\Vert\phi_a  \Vert_2},
\end{aligned}
\end{equation}
where  $\phi_{a}$ is the analytical solution and $\phi_{p}$ is the predicted solution. Meanwhile, the mean-square error  $\text{MSE}$ of the residual of the control equation and the average predicted error $\text{APE}$ of the field are introduced for comparison, which are defined in (\citep{PINN-MDNDE},\cite{jne2040036}), also shown in Eq. \eqref{eq:Case1and2_APEError}.
\begin{equation}\label{eq:Case1and2_APEError}
\begin{aligned}
& \text{MSE}=  \frac{1}{N_{\text{test}}} \sum_{i \in \text{test set}} ( \phi_{p}(r_i) - \phi_a(r_i) )^2\\
&\text{APE}=  \frac{1}{N_{\text{test}}} \sum_{i \in \text{test set}} | \phi_{p}(r_i) - \phi_a(r_i) |
\end{aligned}
\end{equation}
}

\textcolor{black}{For Case-1, the MSE accuracy in \citep{PINN-MDNDE} is 1.4850e-09, and for Case-2, the MSE accuracy in \citep{PINN-MDNDE} is 4.5593e-06, the APE accuracy in \cite{jne2040036} is 1.87e-04. As can be seen from Tab. \ref{tab:Case1and2_Results}, our model outperforms the models reported in \cite{jne2040036,PINN-MDNDE} when solving the simple problems such as Case-1 and Case-2.}


\subsection{Effect of the loss function allocation}
\label{sec:lossfunction}
\textcolor{black}{From this subsection, we will move to 2D IBP and numerous numerical results confirms that, neither the train process of the networks nor the accuracy of the networks is as good as the simple geometry mono-energetic test cases.}

We first discuss the loss function and explore the sensitivity of DEPINN to the weights of different loss terms. Although researchers have proposed many innovative loss functions based on PINN (\citep{Meta-learningPINN}, \citep{lws-PINN}), the most commonly used are MSE and SSE. Both of them minimize the sum of squares of the difference between the value of the target vector and the estimated value. 
The loss function in SSE is used in the present work, due to that it has a larger range to set the hyperparameters in Alg. \ref{algo:1}.

\begin{figure}[htbp]
\centering
\subfloat[ Relative $L_{\infty}$ error of $u$ and $v$]{\includegraphics[width = 0.23\textwidth]{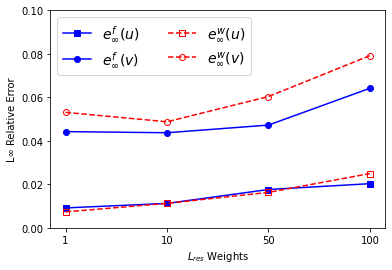}}
\subfloat[ Relative error of $k^{\text{NN}}_\text{eff}$]{\includegraphics[width = 0.23\textwidth]{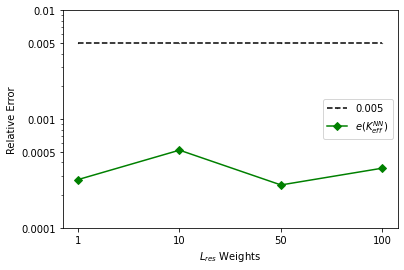}}\\
\subfloat[ Relative $L_{\infty}$ error of $u$ and $v$]{\includegraphics[width = 0.23\textwidth]{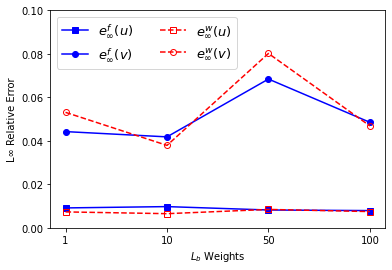}}
\subfloat[ Relative error of $k^{\text{NN}}_\text{eff}$]{\includegraphics[width = 0.23\textwidth]{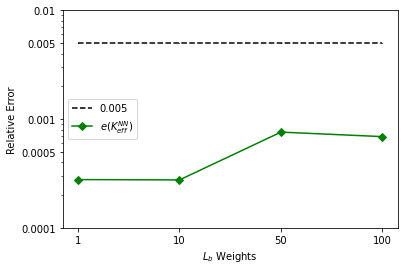}}\\
\subfloat[ Relative $L_{\infty}$ error of $u$ and $v$]{\includegraphics[width = 0.23\textwidth]{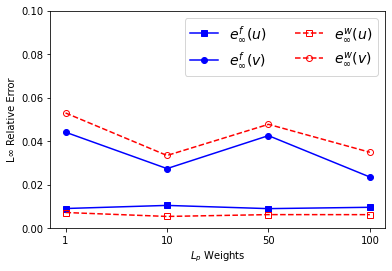}}
\subfloat[ Relative error of $k^{\text{NN}}_\text{eff}$]{\includegraphics[width = 0.23\textwidth]{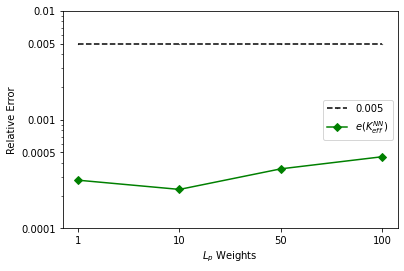}}\\
\caption{(a)-(b) Variations of relative errors of $u$, $v$ and $k^{\text{NN}}_\text{eff}$ when $\alpha_2 =\alpha_3 = 1$ and $\alpha_1$ takes 1, 10, 50, 100 respectively.
(c)-(d) Variations of relative errors of $u$, $v$ and $k^{\text{NN}}_\text{eff}$ when $\alpha_1=\alpha_3 =1$, and $\alpha_2$ takes 1, 10, 50 and 100 respectively. (e)-(f) Variations of relative errors of $u$, $v$ and $k^{\text{NN}}_\text{eff}$ when $\alpha_1 =\alpha_2 =1$ and $\alpha_3$ takes 1, 10, 50 and 100 respectively. (a) (c) (e) Relative errors of $u$ and $v$ in $L_\infty$ norm; (b) (d) (f) Relative errors of $k^{\text{NN}}_\text{eff}$.}
\label{fig:fcsw}
\end{figure}

It has been mentioned that adjusting the weight of the DEPINN loss function term will effectively improve the results (\citep{PINNTA1}, \citep{Weighted-Loss}). In the following experiment (Fig. \ref{fig:fcsw}), we set different values for $\alpha_1$, $\alpha_2$ and $\alpha_3$ in Eq. \eqref{eq:PINNloss} to show how they affect the neural networks.

In Fig. \ref{fig:fcsw}, due to the different weights of the loss function, the value of the loss function varies greatly. 
We take the optimization epochs of Adam algorithm and LBFGS algorithm to be 50000 epochs. 
From Fig. \ref{fig:fcsw}(a)-(d), since the number of prior points $\bx_p$ is very small,
$\mathcal{L}_{res}$ and $\mathcal{L}_{b}$ contribute a large proportion of the total loss function; when  the weights of $\mathcal{L}_{res}$ and $\mathcal{L}_{b}$ are increased, it is very difficult to obtain a good result, and even worse. From Fig. \ref{fig:fcsw}(e)-(f), when the weight of $\alpha_3$ of $L_{p}$ increases, the relative error decreases. When $\alpha_3$ is increased to 10, the result is the best one in experiment \ref{fig:fcsw}. Therefore, we choose $ \alpha_1=1, \alpha_2=1$ and $\alpha_3=10$ for the following experiments.

\subsection{Effect of the optimization strategy}
\label{sec:optimization}
In this section, we compare the effects of various training strategies and explore the sensitivity of the hyperparameters $S_1$ and $\epsilon_1$ in Alg. \ref{algo:1}. There are many optimization methods for neural networks. Many researchers have tried to combine PINN with other optimization algorithms (\citep{PIELM}, \citep{DTPINN}), but the most widely used algorithm is Adam. In order to demonstrate the effectiveness of the adaptive training strategy proposed in this article, the following experiments are compared with two other training strategies, i.e., Adam and LBFGS optimization methods. 

\begin{itemize}
\item[(i)] Strategy-1: Fix 5000 training epochs of Adam.
\item [(ii)]Strategy-2: Stop Adam training when the value of the loss function is less than a threshold (we choose 15).
\item[(iii)] Strategy-3: Adaptive method (Alg. \ref{algo:1}).
\end{itemize}
The stopping criteria of LBFGS algorithms of these three strategies are all set as: if the loss function is less than a certain threshold, the training stops.
\begin{figure}[htbp]
\centering
\subfloat[Relative $L_{\infty}$ error of $u$ and $v$]{\includegraphics[width = 0.23\textwidth]{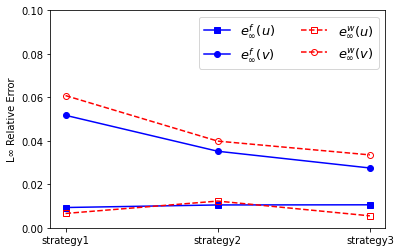}}
\subfloat[Total Loss $\mathcal{L}$ during training process]{\includegraphics[width = 0.23\textwidth]{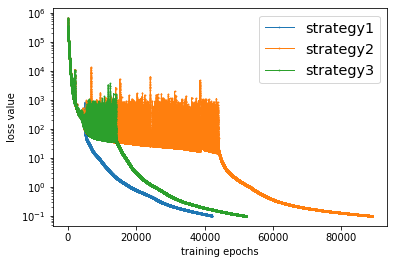}}
\caption{Variations of relative errors of $u$, $v$ ,$ k^{\text{NN}}_\text{eff}$ and the total Loss $\mathcal{L}(\bx,\lambda;W)$ with three training strategies. (a) Relative error of $u$ and $v$ in $L_\infty$ norm; (b) Total Loss $\mathcal{L}(\bx,\lambda;W)$ during the whole optimization procedure.}
\label{fig:fcts}
\end{figure}

In Fig. \ref{fig:fcts}, the performance of Strategy-1 shows that less Adam training may lead to worse results, and more LBFGS training is required to satisfy the stopping criterion. From the results of Strategy-2, training Adam too many epochs is inefficient, so early stopping is a good choice. Actually, Strategy-2 can avoid the problem that the loss function fluctuates to a larger value when starting LBFGS optimization, but it depends on the threshold setting strictly.
In contrast, Strategy-3 has more advantages, which can be observed in Fig. \ref{fig:fcts}.


\begin{figure}[htbp]
\centering
\subfloat[Relative $L_{\infty}$ error of $u$ and $v$]{\includegraphics[width = 0.23\textwidth]{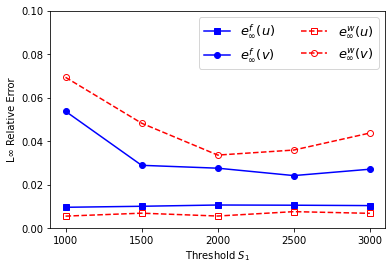}}
\subfloat[Total Loss $\mathcal{L}$ during training process]{\includegraphics[width = 0.23\textwidth]{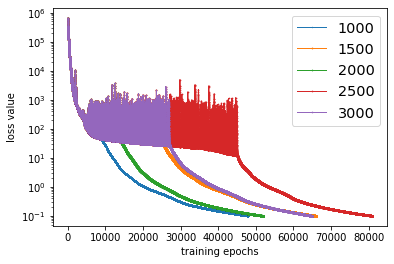}}\\
\subfloat[Relative $L_{\infty}$ error of $u$ and $v$]{\includegraphics[width = 0.23\textwidth]{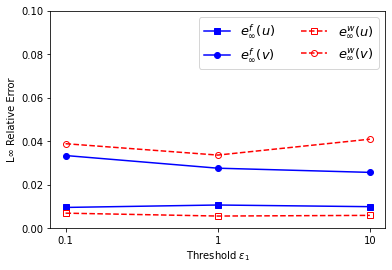}}
\subfloat[Total Loss $\mathcal{L}$ during training process]{\includegraphics[width = 0.23\textwidth]{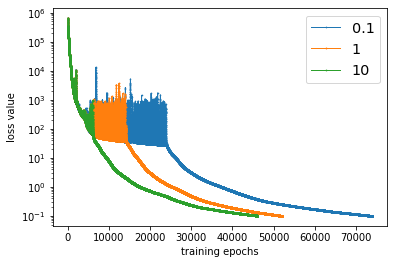}}
\caption{(a)-(b) Variations of relative errors of $u$, $v$, the total losses $\mathcal{L}(\bx,\lambda;W)$ when $\epsilon_1 = 1.0$ in Algorithm \ref{algo:1} and $S_1$ takes 1000, 1500, 2000, 2500 and 3000 respectively. (c)-(d) Variations of relative errors of $u$, $v$, the total Losses $\mathcal{L}(\bx,\lambda;W)$ when $S_1 = 2000$ in Algorithm \ref{algo:1} and $\epsilon_1$ takes 0.1, 1 and 10 respectively. (a) (c) Relative errors of $u$ and $v$ in $L_\infty$ norm; (b) (d) Total losses $\mathcal{L}(\bx,\lambda;W)$.}
\label{fig:S1S2_Set}
\end{figure}

It is observed from Fig. \ref{fig:S1S2_Set}(a)-(b) that when $S_1 = 2000$, taking into account the error and the number of training times, the result is the best. Besides, if $S_1$ is set too smaller, the difference average $\overline{\text{dif}}_1$ and $\overline{\text{dif}}_2$ in Algorithm \ref{algo:1} is more likely to be too close, and the Adam will end too earlier which might lead to poor results. As shown in Fig. \ref{fig:S1S2_Set}(c)-(d), when $S_1$ is fixed, the Adam ends faster as $\epsilon_1$ is larger. Considering the error and the number of training epochs, when $\epsilon_1 = 1.0$, the result is the best. In order to fix the parameters, we choose $ S_1=2000, \epsilon_1=1.0 $ in subsequent experiments.

\subsection{Effect of the sampling procedure}
\label{sec:sampling}

\textcolor{black}{In this section, we explore how to choose the prior points and the sensitivity of the neural network to the sample rates. In the process of solving the 2D IBP with PINN, there are three types of points: point $\bx_r$ in the computational domain, point $\bx_b$ on the boundary and point $\bx_p$ corresponds to prior solution. Note that these two types of points $\bx_r$ and $\bx_b$ have a large number (hundreds and thousands), but the number of $\bx_p$ is only 38 × 2. Although the number of $x_p$ is very small, it has a great impact on the training and prediction of neural networks(See TABLE \ref{tab:DLT_Benchmark} and \ref{tab:PredictComparison}). Here, other researchers also have some research on how to select prior data \cite{Sampling1}, \cite{Sampling2} and \cite{Sampling3}, which can be read by interested readers. Since the number of $\bx_p$ is small, we decided to design an experiment to explore its sampling strategy in detail.}


Since the physical coefficients of Eq. \eqref{eq:diffusion} in $\Omega_1$ and $\Omega_4$ are quite different, we introduce a strategy on how to choose the prior points distribution: 
\begin{itemize}
\item[(i)] Fix the prior points of $\Omega_{1,2,3}$, the prior points of $\Omega_4$ are gradually moved from the junction of $\Omega_1$ and $\Omega_4$ to the center (2cm each time).
\item[(ii)] Fix the prior points of $\Omega_4$ with the best result.
\item[(iii)] Follow the same procedure in (i) and (ii) to fix the prior points of $\Omega_1$.
\end{itemize}



\begin{figure}[htbp]
\centering
\subfloat[ Relative $L_{\infty}$ error of $u$ and $v$]{\includegraphics[width = 0.23\textwidth]{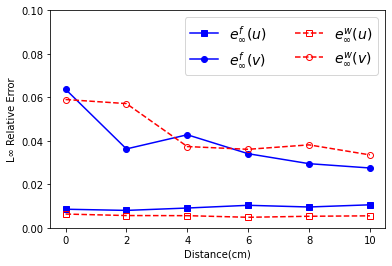}}
\subfloat[Relative error of $ k^{\text{NN}}_\text{eff}$]{\includegraphics[width = 0.23\textwidth]{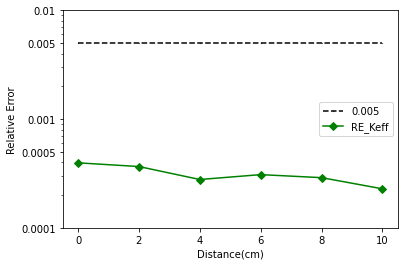}}\\
\subfloat[ Relative $L_{\infty}$ error of $u$ and $v$]{\includegraphics[width = 0.23\textwidth]{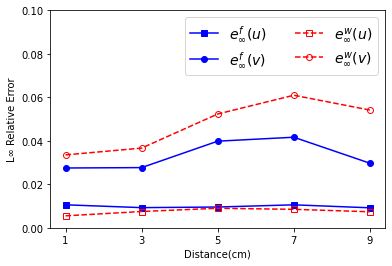}}
\subfloat[Relative error of $ k^{\text{NN}}_\text{eff}$]{\includegraphics[width = 0.23\textwidth]{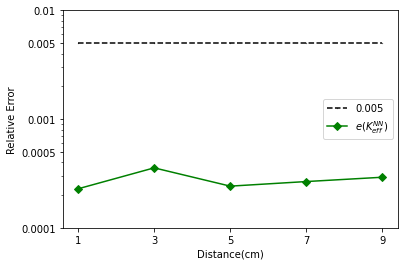}}
\caption{(a)-(b) Variations of relative errors of $u$, $v$ and $ k^{\text{NN}}_\text{eff}$ when the prior points of $\Omega_{1,2,3}$ are fixed, the distance between the prior points of $\Omega_4$ and the junction of $\Omega_1$ and $\Omega_4$ is 0, 2, 4, 6, 8, 10 respectively. (c)-(d) Variations of relative errors of $u$, $v$ and $ k^{\text{NN}}_\text{eff}$ when the prior points of $\Omega_{2,3,4}$ are fixed, the distance between the prior points of $\Omega_1$ and the junction of $\Omega_1$ and $\Omega_4$ is 1, 3, 5, 7, 9 respectively. (a) (c) Relative errors of $u$ and $v$ in $L_\infty$ norm; (b) (d) Relative errors of $ k^{\text{NN}}_\text{eff}$.}
\label{fig:fcdp_r4r1}
\end{figure}

Since the prior points are symmetric about $y = x$, only the prior points in the lower half of the area are drawn in Fig. \ref{fig:iaeacore}. Most of the prior points (black points) in the $\Omega_2$ and $\Omega_3$ are fixed at the center of each small region, the prior points (pink stars) of $\Omega_1$ are fixed on the junction of $\Omega_1$ and $\Omega_4$.  Fig. \ref{fig:fcdp_r4r1}(a)-(b) show the situation when prior points (red rhombuses) in $\Omega_4$ are fixed at the junction, and the remaining seven points (orange rhombuses) gradually move from the junction to the center at the rate of 2cm per experiment. After fixing the prior points of of $\Omega_2$, $\Omega_3$ and $\Omega_4$, Fig. \ref{fig:fcdp_r4r1}(c)-(d) show the situation when the prior points (pink stars) of $\Omega_1$ gradually moves from the junction to the center (2cm each time).

From Fig. \ref{fig:fcdp_r4r1}, it is observed that when the distance of the prior points of $\Omega_4$ from the junction of $\Omega_1$ and $\Omega_4$ increases, the results become better, and when the distance of the prior points of $\Omega_1$ from the junction increases, the results get worse.

From Tab. \ref{tab:coefs}, the coefficients change at the junction is more severely than that in the interior of $\Omega_1$, so fixing the prior points of $\Omega_1$ at the junction is more conducive to the neural networks learning of the flux of $\Omega_1$ and $\Omega_4$. But in $\Omega_4$, the fluxes of the whole region change severely. Therefore, in the following experiments, the prior points will be sampled according to the configuration of Fig. \ref{fig:iaeacore}.

\begin{figure}[htbp]
\centering
\subfloat[ Relative $L_{\infty}$ errors of $u$ and $v$]{\includegraphics[width = 0.23\textwidth]{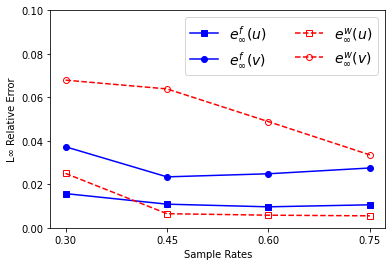}}
\subfloat[Relative error of $ k^{\text{NN}}_\text{eff}$]{\includegraphics[width = 0.23\textwidth]{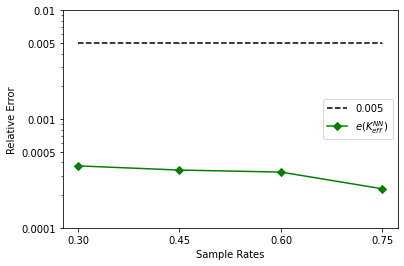}}
\caption{Variations of relative errors of $u$, $v$ and $ k^{\text{NN}}_\text{eff}$ when the sampling rates of $x_r$ and $x_b$ take 0.30, 0.45, 0.60 and 0.75 respectively. (a) Relative error of $u$ and $v$ in $L_\infty$ norm; (b) Relative error of $k^{\text{NN}}_\text{eff}$.}
\label{fig:fcsr}
\end{figure}

Now we explore the sensitivity of the neural networks to the sample rates of $\bx_r$ and $\bx_b$.
By the uniform grid distribution used in test data, we can obtain nearly 20,000 discrete grid points in the entire computing domain. We use random sampling under an uniform distribution to sample these discrete grid points at a ratio, which is the sample rate. 
Numerical results obtained are shown in Fig. \ref{fig:fcsr}, when the sample rates of $x_r$ and $x_b$ take 0.30, 0.45, 0.60 and 0.75 respectively.

According to experimental observations, when DEPINN is used to solve the 2D IBP, the simulation time is almost proportional to the sample rates. According to Fig. \ref{fig:fcsr}, the results become better when the sampling rate increases, which is consistent with the usual cognition about neural networks. Considering the sampling rates of 0.60 and 0.75, the difference of simulation time is not significant. Therefore, we choose the sampling rate as 0.75 in the subsequent experiments.

\begin{figure}[htbp]
\centering
\subfloat[ FreeFem++ results of $u$]{\includegraphics[width = 0.25\textwidth]{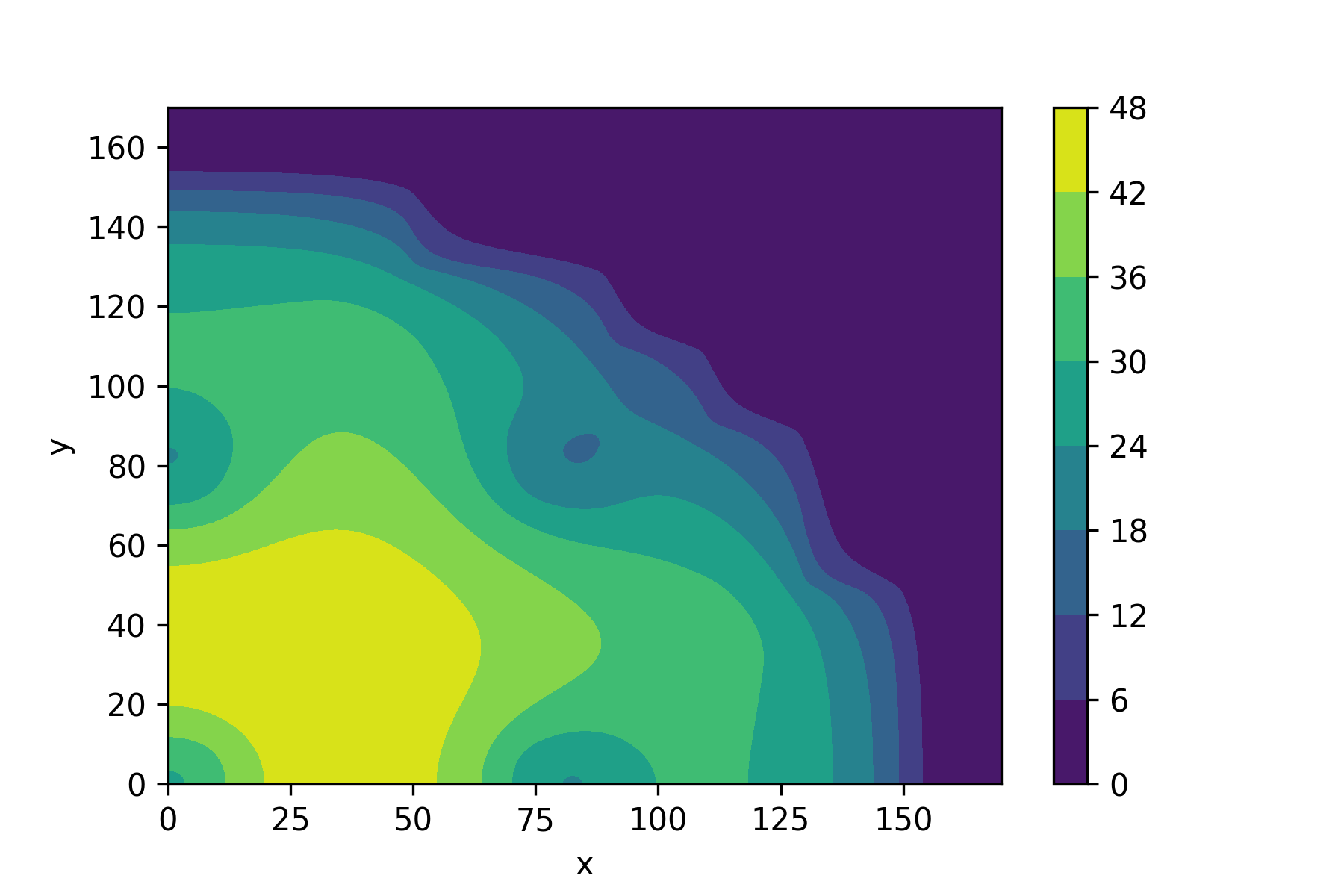}}
\subfloat[ FreeFem++ results of $v$]{\includegraphics[width = 0.25\textwidth]{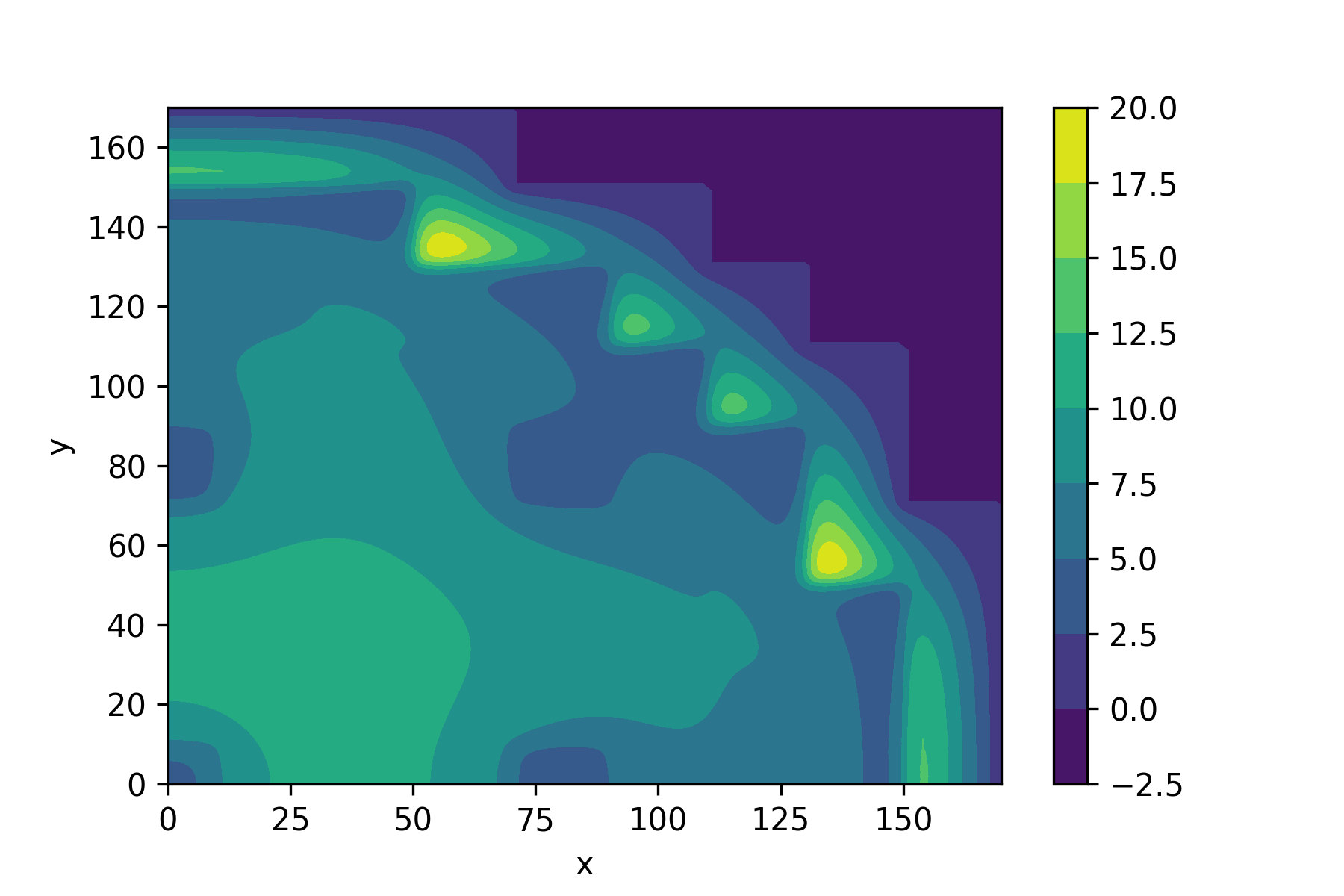}}\\
\subfloat[ PINN results of $u$]{\includegraphics[width = 0.25\textwidth]{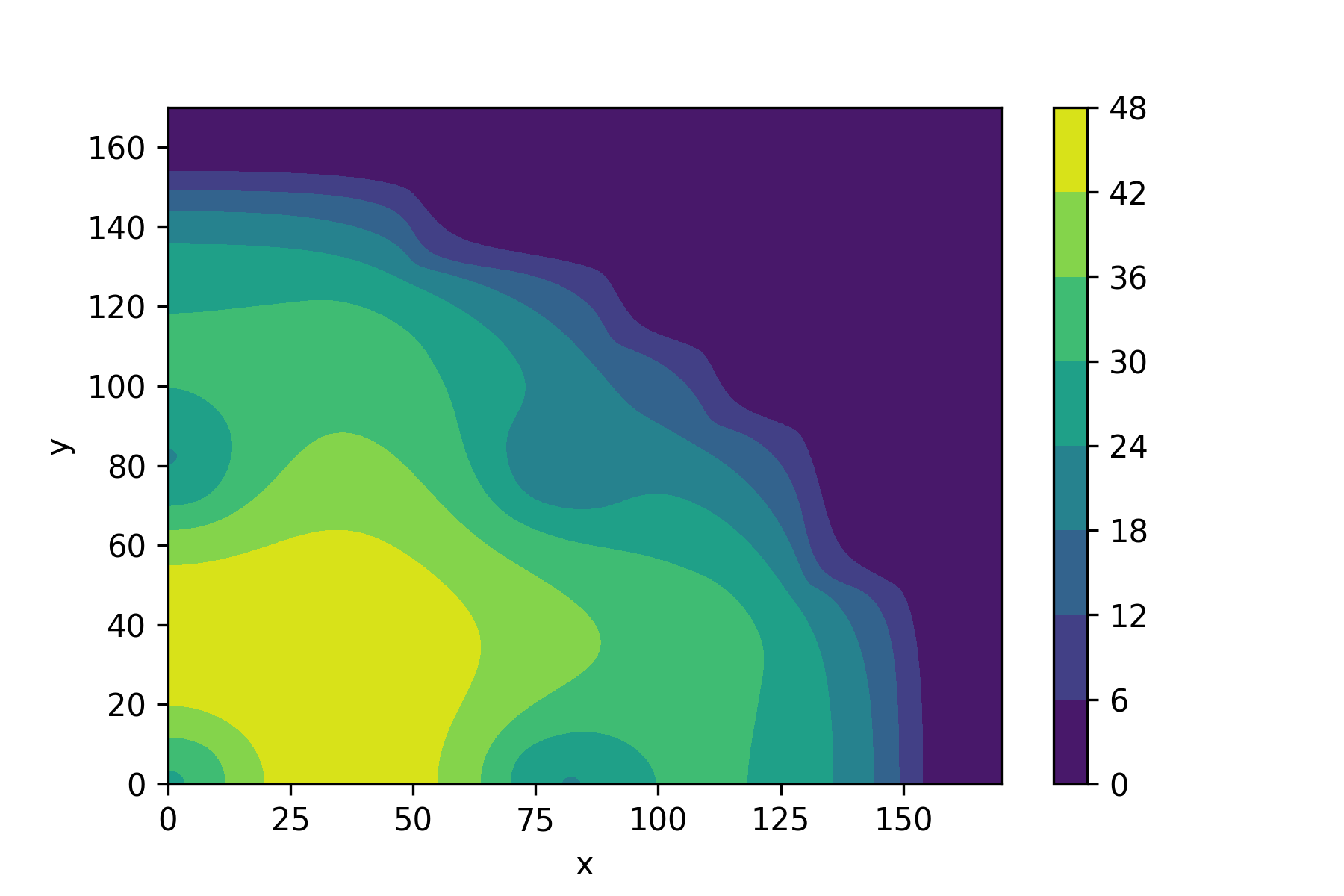}}
\subfloat[ PINN results of $v$]{\includegraphics[width = 0.25\textwidth]{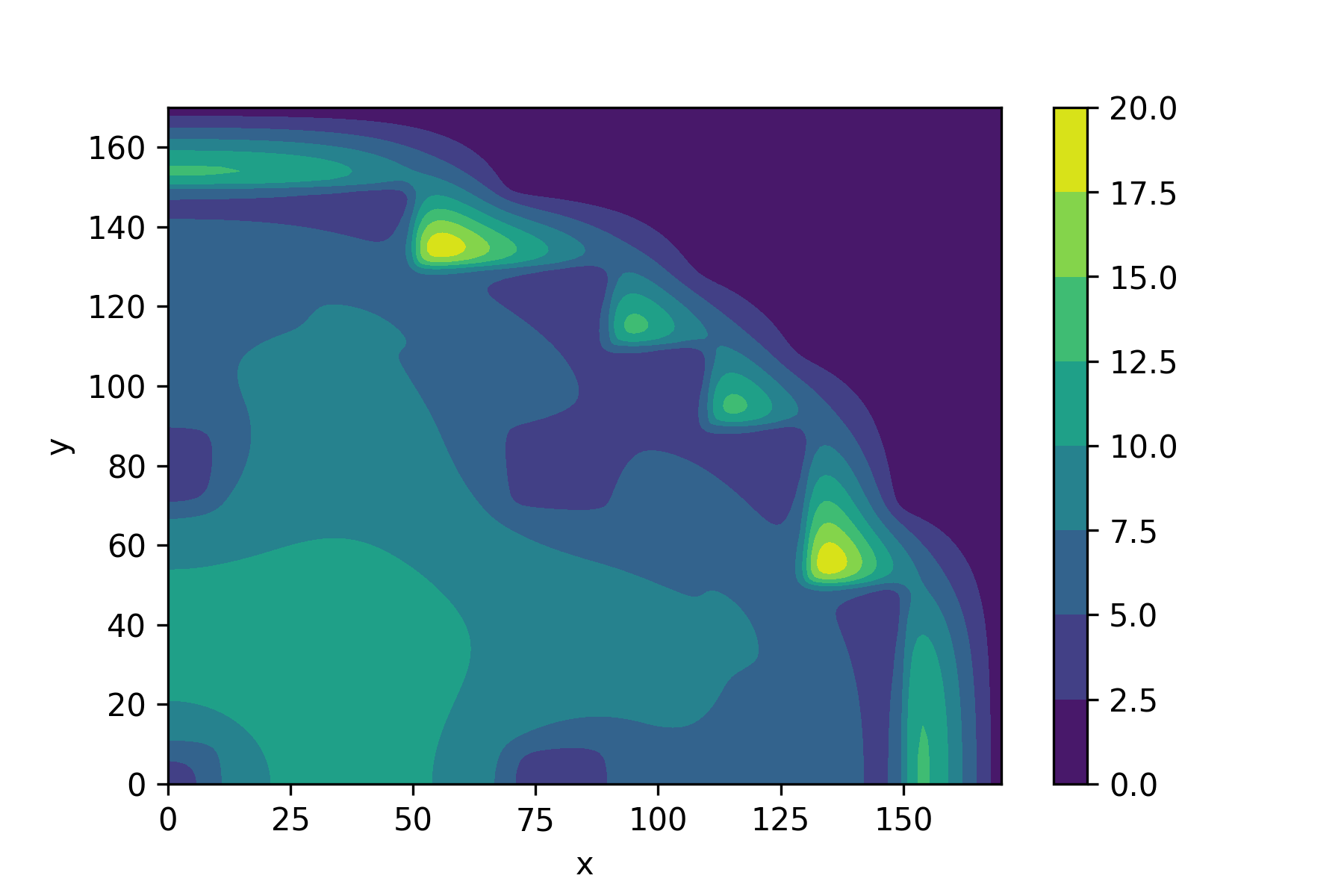}}\\
\subfloat[ Absolute error of $u$]{\includegraphics[width = 0.25\textwidth]{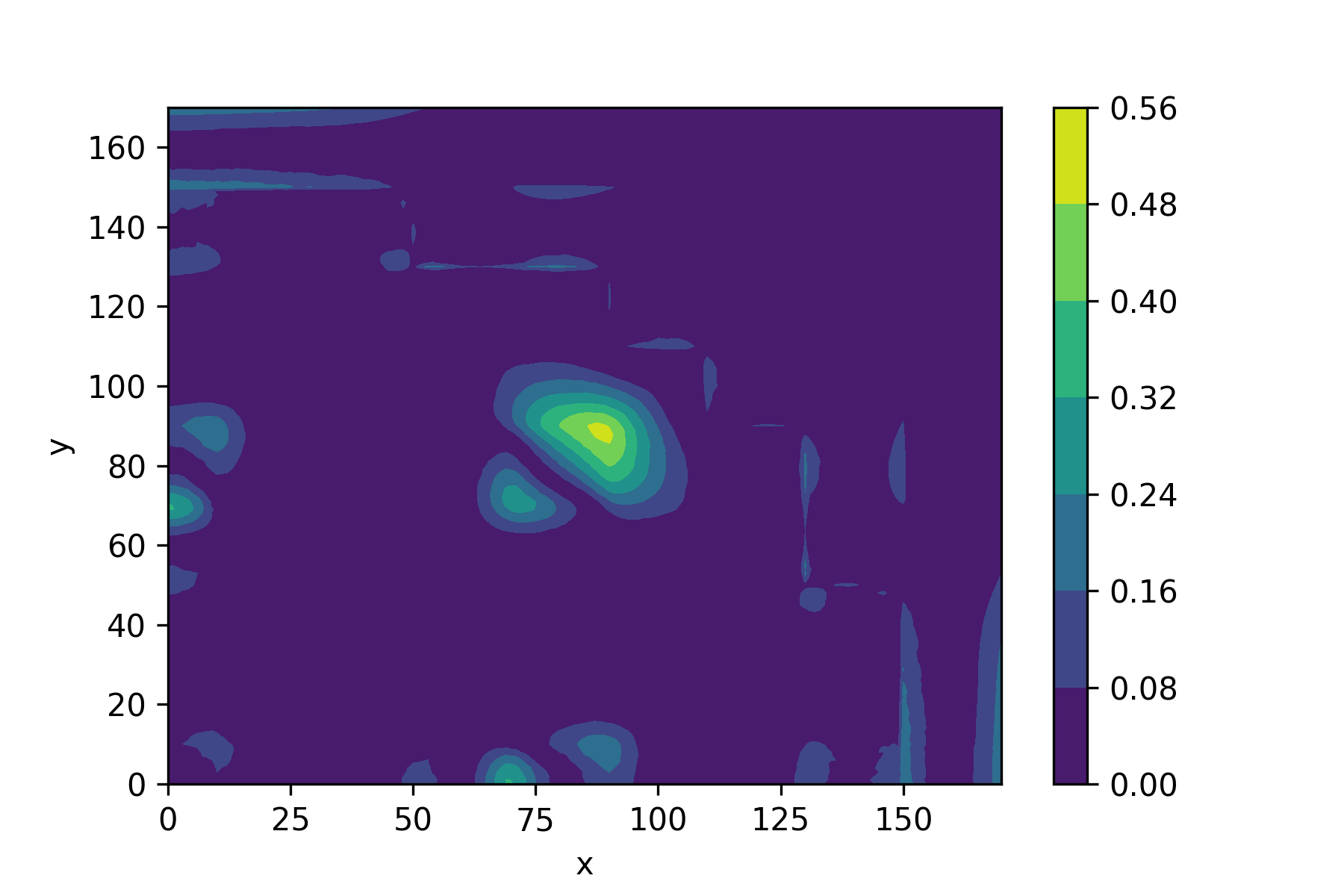}}
\subfloat[ Absolute error of $v$]{\includegraphics[width = 0.25\textwidth]{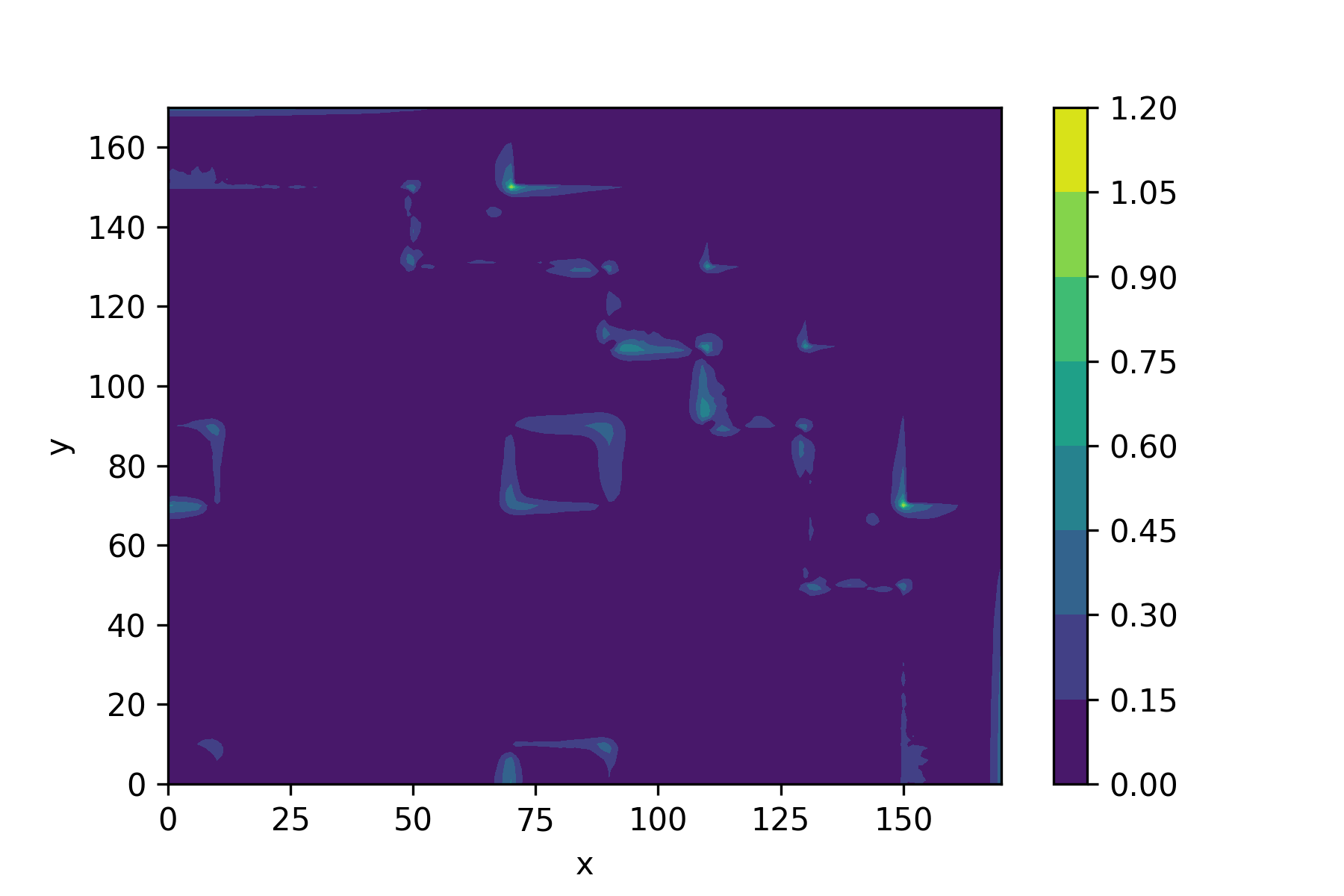}}\\
\caption{After a large number of experiments to solidify the parameters, the DEPINN model in this paper is compared with the commonly used finite element calculation software FreeFem++ on 2D IBP. (a)-(b) illustrate the result of $u$ and $v$ predicted by FreeFem++. (c)-(d) illustrate the result of $u$ and $v$ predicted by DEPINN. (e)-(f) present the absolute error (defined in Fig. \ref{fig:Case1and2}(d)) of $u$ and $v$. We find that relative large errors appear around the corner and the interface of different materials.}
\label{fig:IAEA_Heatmap}
\end{figure}

To this end, we show in Fig. \ref{fig:IAEA_Heatmap} the comparison of the DEPINN model and the Freefem++ model on 2D IBP, where the latter we take it as reference. For the DEPINN model, all the parameter settings are optimal benefit from the learning procedure of previous subsections. In this figure, we illustrate in (a)-(b) the result of $u$ and $v$ predicted by FreeFem++; (c)-(d) the result of $u$ and $v$ predicted by PINN; (e)-(f) the absolute error (defined in Fig. \ref{fig:Case1and2}(d)) of $u$ and $v$. We find that relative large errors appear around the corner and the interface of different materials, thus more training points or prior data are necessary for further improving the accuracy of the DEPINN model. This find is consistent with the result in Section \ref{sec:sampling}. We emphasize that the 2D IBP is much more complex than test Case-1 and Case-2, thus the accuracy of our DEPINN degenerates, but still satisfies the engineering acceptance criteria. Indeed, the 2D IBP is adapted from real engineering problem \citep{Benchmark}, which confirms again that the proposed DEPINN has potential for real engineering applications.

\subsection{Performance on parameter dependence}
\label{sec:parameter}

In this section, we investigate the performance on parameters dependence on the proposed architecture. After fixing some parameters of DEPINN according to the above sections, we need to check the generalization performance of the model. When changing the coefficients in Tab. \ref{tab:coefs} of the 2D IBP, specifically, increasing $\Sigma_{a,2}$ of $\Omega_2$ and $\Omega_3$ from 0.080 to 0.130 with step size 0.005, we explore the performance of existing model in these 22 experiments with various coefficients.

\begin{table}[h]
\centering
\caption{The variation of $\be_{\infty}^{\text{f}}(u)$ , $\be_{\infty}^{\text{f}}(v)$ and $\be(k^{\text{NN}}_{\text{eff}})$ when  $\Sigma_{a,2}$ of $\Omega_2$ and $\Omega_3$ are taken from 0.080 to 0.130.}
\setlength{\tabcolsep}{0.2mm}{
\begin{tabular}{c|cccccc}
\hline\noalign{\smallskip}
$\Sigma_{a,2}$ & $\be_{\infty}^{\text{f}}(u)_{\Omega_{2}}$ & $\be_{\infty}^{\text{f}}(u)_{\Omega_{3}}$ & $\be_{\infty}^{\text{f}}(v)_{\Omega_{2}}$  & $\be_{\infty}^{\text{f}}(v)_{\Omega_{3}}$ & $\be(k^{\text{NN}}_{\text{eff}})_{\Omega_{2}}$ & $\be(k^{\text{NN}}_{\text{eff}})_{\Omega_{3}}$ \\
\hline
0.080 & 0.0061 & 0.0030 & 0.0301 & 0.0317 & 0.0001 & 0.0001 \\
0.085 & 0.0105 & 0.0027 &  0.0275 & 0.0243 & 0.0002 & 0.0001 \\
0.090&	0.0136&	0.0068&	0.0327&	0.0452&	0.0010&	0.0002\\
0.095&	0.0239&	0.0064	&0.0295&	0.0403	&0.0010&	0.0003\\
0.100	&0.0232	&0.0051	&0.0363	&0.0512	&0.0040	&0.0003\\
0.105&	0.0153&	0.0082&	0.0499&	0.0357&	0.0049&	0.0004\\
0.110&	0.0214&	0.0077&	0.0453&	0.0334&	0.0040&	0.0002\\
0.115&	0.0244&	0.0068&	0.0348&	0.0430&	0.0033&	0.0003\\
0.120&	0.0299&	0.0091&	0.0487&	0.0348&	0.0039&	0.0001\\
0.125&	0.0275&	0.0134&	0.0451&	0.0398&	0.0016&	0.0002\\
0.130&	0.0289&	0.0105&	0.0495&	0.0275&	0.0038&	0.0002\\
\hline
\end{tabular}
}
\label{tab:fcdc_r2} 
\end{table}


\begin{table}[h]
\centering
\caption{
The results meet the acceptance criteria.
}
\setlength{\tabcolsep}{1.mm}{
\begin{tabular}{c|ccccc}
\hline\noalign{\smallskip}
Variables & $\operatorname*{max}\limits_{<0.9} $   & $\operatorname*{max}\limits_{<0.9}$ & $\operatorname*{max}\limits_{>0.9}$   &$\operatorname*{max}\limits_{>0.9}$ &$\operatorname*{max}$ \\
& $\be_{\infty}^{\text{f}}(u)$   & $\be_{\infty}^{\text{f}}(v)$ & $\be_{\infty}^{\text{f}}(u)$   &$\be_{\infty}^{\text{f}}(v)$ &$\be(k^{\text{NN}}_{\text{eff}})$ \\
\hline
$\Omega_2, \Sigma_{a,2}$ & 0.0016 & 0.0267 & 0.0299 & 0.0498 & 0.0049\\
\hline
$\Omega_3, \Sigma_{a,2}$  & 0.0026 & 0.0512  & 0.0134 & 0.0398 & 0.0004\\
\hline
\end{tabular}
}
\label{tab:fcdc_max} 
\end{table}

Tab. \ref{tab:fcdc_r2} shows the relative errors with respect to the variation of $\Sigma_{a,2}$ in region $\Omega_2$ and $\Omega_3$. When the relative flux is less than or greater than 0.9, the maximum value of the relative error to determine whether it satisfies the acceptance criteria  is shown in Tab. \ref{tab:fcdc_max}.
All the relative errors of $u, v$ and $k^{\text{NN}}_\text{eff}$ satisfy the acceptance criteria (Section \ref{sec:measures}), which fully illustrate the generalization performance of our DEPINN model. We conclude here that the DEPINN model we proposed in this manuscript can be used for engineering scale problems in nuclear reactor physics.

\begin{table}[h]
\centering
\caption{
\textcolor{black}{Summary of the accuracy of the field solved with PINN for three test cases.}
}
\setlength{\tabcolsep}{0.2mm}{
\begin{tabular}{c|c|c|c|c|c|c}
\hline\noalign{\smallskip}
& $\be_{\infty}(\phi_p)$ & $\be_{2}(\phi_p)$ & $\be_{\infty}(u)$  & $\be_{\infty}(v)$ & $\be_{2}(u)$ & $\be_{2}(v)$ \\
\hline
Case-1 & 3.7615e-05 & 1.7018e-05 &  &  &  &  \\
Case-2 & 3.6901e-04 & 1.2004e-04 &   &  &  &  \\
2D IBP $\Omega_1$&	&	&	0.0103 & 0.0275 & 0.0048&	0.0125\\
2D IBP $\Omega_2$&	&	&  0.0078&	0.0243	&0.0013&	0.0044\\
2D IBP $\Omega_3$&  &	& 0.0105 & 0.0254	&0.0077	&0.0315\\
2D IBP $\Omega_4$&	&	&	0.0055&	0.0335&	0.0153&	0.0148\\

\hline
\end{tabular}
}
\label{tab:ForComparison} 
\end{table}



\section{Conclusion}
\label{sec:conclusion}
\textcolor{black}{In this work, a data-enabled physics-informed neural network (DEPINN) is proposed for solving neutron diffusion eigenvalue problems (NDEPs).   The feasibility of proposed DEPINN model is tested on three typical benchmark problems, from simple geometry to complex geometry, and from mono-energetic equation to two-group equations.  Particularly, a comprehensive numerical study of DEPINN for solving engineering scale NDEPs i.e. IAEA benchmark problem is presented. 
}

\textcolor{black}{Different from simple geometry and analytical cases where high accuracy is achieved, the train procedure for the engineering scale problem is rather complicated. \textcolor{black}{In order to achieve an engineering acceptable accuracy, a very small amount of prior data are suggested to be used to improve the accuracy and efficiency of training, where we assume the data can be obtained from engineering experiments.} To improve the efficiency of the network, an adaptive optimization procedure with Adam and LBFGS is proposed to accelerate the convergence in the training stage. Furthermore, we discuss the effect of different physical parameters, sampling techniques, loss function allocation and the generalization performance of the proposed DEPINN model for solving complex problems.
}

\textcolor{black}{Numerical results based on three test cases show that the proposed DEPINN model can efficiently solve the neutron diffusion eigenvalue equations. Based on the test on IAEA benchmark problem, we conclude that the proposed framework with DEPINN is possible for real nuclear engineering problems in nuclear reactor physics domain, particularly where observations are available. Further works shall be brought such as i) test on three-dimensional, large-scale reactor physics problems, ii) performance on noise observations and iii) new PINN structure for real complex engineering problem where no observation is available, etc.
}

\section*{Contribution statement}
\noindent Yu Yang: Methodology, Coding,  Writing \& Editing. Helin Gong: Conceptualization, Methodology, Nuclear engineering data curation, Writing \& Editing, Review, Funding acquisition. Shiquan Zhang: Conceptualization, Methodology, Review, Funding acquisition. Qihong Yang: Methodology, Coding. Zhang Chen: Funding acquisition, Validation. Qiaolin He: Conceptualization, Methodology, Review, Funding acquisition. Qing Li: Supervision, Review.


\section*{Acknowledgment}

This research is supported part by the National Natural Science Foundation of China (No.11971020, 11905216, 12175220), and the Stability Support Fund for Science and Technology on Reactor System Design Technology Laboratory. \textcolor{black}{The authors are grateful to the three anonymous reviewers’ constructive suggestions for the work during the preparation of the manuscript. 
All data and codes used in this manuscript are publicly available on GitHub at https://github.com/YangYuSCU/DE-PINN.
} 


%





\ifCLASSOPTIONcaptionsoff
\newpage
\fi




\bibliographystyle{IEEEtran}
\bibliography{IAEAPINN}

\vfill


\end{document}